\documentclass[particles,article,accept,oneauthor,pdftex]{Definitions/mdpi} 

\firstpage{1095} 
\pubvolume{7}
\issuenum{4}
\articlenumber{67}
\pubyear{2024}
\copyrightyear{2024}
\externaleditor{Academic Editor: Armen Sedrakian}
\datereceived{20 September 2024} 
\daterevised{7 November 2024}
\dateaccepted{22 November 2024} 
\datepublished{12 December 2024} 
\hreflink{https://\linebreak doi.org/10.3390/particles7040067} 

\Title{On the exact solution for the Schrödinger equation}

\TitleCitation{On the exact solution for the Schrödinger equation}

\Author{Yair Mulian $^{1,2,3,4}$}

\AuthorNames{Yair Mulian}

\AuthorCitation{Mulian, Y.}

\address{%
$^{1}$ \quad Institute of Modern Physics, Chinese Academy of Sciences, Lanzhou 730000, China\\
$^{2}$ \quad School of Nuclear Physics, University of Chinese Academy of Sciences, Beijing 100049, China\\
$^{3}$ \quad CAS Key Laboratory of High Precision Nuclear Spectroscopy, Institute of Modern Physics, Chinese Academy of Sciences, Lanzhou 730000,
 China\\
$^{4}$ \quad Instituto Galego de Fisica de Altas Enerxias IGFAE, Universidade de Santiago de Compostela,\linebreak  15782 Santiago de Compostela, Spain\\Email: yair25m@gmail.com}
\corres{Correspondence: }

\abstract{For almost 75 years, the general solution for the Schrödinger equation was assumed to be generated by an exponential or a time-ordered exponential known as the Dyson series. We study the unitarity of a solution in the case of a singular Hamiltonian and provide a new methodology that is not based on the assumption that the underlying space is $L^{2}(\mathbb{R})$. Then, an alternative operator for generating the time evolution that is manifestly unitary is suggested, regardless of the choice of Hamiltonian. The new construction involves an additional positive operator that normalizes the wave function locally and allows us to preserve unitarity, even when dealing with infinite dimensional or non-normed spaces. Our considerations show that Schrödinger and Liouville equations are, in fact, two sides of the same coin and together they provide a unified description for unbounded quantum~systems.}


\begin{document}
\section{Introduction}
The Schrödinger equation (named after Erwin Schrödinger, who postulated the equation in 1925) is the fundamental operatorial equation that governs the wave function of a quantum mechanical system~\cite{Schro}. The discovery of this linear differential equation was a significant landmark in the development of quantum mechanics. In basic terms, if a wave function of a system $\left|\Psi(t_{0}\right\rangle$ is known at some moment $t_{0}$, one can determine the wave function at all the subsequent moments by solving the equation (adopting the convention $\hbar=1$) \cite{landau1,landau2,landau3,landau4,landau5,Peskin1,Peskin2,Peskin3,Peskin4}:
\begin{equation}\label{tindep}
\frac{d}{dt}\left|\Psi(t)\right\rangle \,=\,-i\hat{H}\left|\Psi(t)\right\rangle .
\end{equation}

As long as the \textls[-25]{Hamiltonian of the system $\hat{H}$ is expressible as a bounded time independent operator (a linear operator $\mathcal{\hat{O}}:\mathcal{D}(\mathcal{\hat{O}})\rightarrow\mathcal{Y}$ is called bounded only if $\sup_{||\left|\psi\right\rangle ||=1}\|\mathcal{\hat{O}}\left|\psi\right\rangle \|_{\mathcal{Y}}<\infty$} for any $\left|\psi\right\rangle \in\mathcal{D}(\mathcal{\hat{O}})$ \cite{func1,func2,fu1,fu2}), the solution of (\ref{tindep}) is given by the unitary evolution,
\begin{equation}\label{Udefsimple}
\left|\Psi(t)\right\rangle \,=\,\hat{U}(t,\,t_{0})\left|\Psi(t_{0})\right\rangle ,\qquad\qquad\hat{U}(t,\,t_{0})=\exp\left[-i\int_{t_{0}}^{t}dt^{\prime}\,\hat{H}\right].
\end{equation}

In various cases the Hamiltonian of a quantum system includes a time-dependent part, and Equation (\ref{tindep}) is replaced by
\begin{equation}\label{schrodtime}
\frac{d}{dt}\left|\Psi(t)\right\rangle \,=\,-i\hat{H}(t)\left|\Psi(t)\right\rangle ,\qquad or\qquad\frac{d\hat{U}(t,\,t_{0})}{dt}\,=\,-i\hat{H}(t)\,\hat{U}(t,\,t_{0}).
\end{equation}

\textls[-40]{The known general solution for (\ref{schrodtime}) is expressible by using the time ordering $\hat{T}$ operator~\cite{freema1,freema2},}
\begin{equation}\label{Udef}
\hat{U}(t,\,t_{0})\,=\,\hat{T}\exp\left[-i\int_{t_{0}}^{t}dt^{\prime}\,\hat{H}(t^{\prime})\right],
\end{equation}
where the action of the $\hat{T}$ operator is defined as
\begin{equation}
\hat{T}\left[\hat{H}(t^{\prime})\,\hat{H}(t)\right]\,\equiv\,\Theta(t^{\prime}-t)\hat{H}(t^{\prime})\,\hat{H}(t)\,+\,\Theta(t-t^{\prime})\hat{H}(t)\,\hat{H}(t^{\prime}),
\end{equation} 
with $\Theta(t^{\prime}-t)$ denoting the Heaviside theta function.
 
 \indent It is well known that solutions (\ref{Udefsimple}) and (\ref{Udef}) are unitary operators on Hilbert spaces for any choice of self-adjoint Hamiltonians. However, as we shall discuss, the property of unitarity no longer holds on more general choice of spaces, unless the Hamiltonian is a strictly bounded self-adjoint operator.
 Our claim in this paper is that instead of \mbox{solutions (\ref{Udefsimple}) and (\ref{Udef})}, one can obtain a universal unitary solution (our solution is denoted by $\mathcal{\hat{P}}$ as a shortening to the word \textit{pitaron}, meaning solution in Hebrew):
\begin{equation}\label{pitaron}
\left|\Psi(t)\right\rangle \,=\,\mathcal{\hat{P}}(t,\,t_{0})\left|\Psi(t_{0})\right\rangle \,,\qquad\qquad\mathcal{\hat{P}}(t,\,t_{0})\,\equiv\,\hat{\mathcal{N}}(t,\,t_{0})\,\hat{U}(t,\,t_{0}),
\end{equation}
with the new operator,
\begin{equation}\label{definorm}
\mathcal{\hat{N}}(t,\,t_{0})\,\equiv\,\sqrt{\hat{U}^{\dagger-1}(t,\,t_{0})\,\hat{U}^{-1}(t,\,t_{0})}\,.
\end{equation} 

The new operator $\mathcal{\hat{N}}(t,\,t_{0})$ from (\ref{definorm}) is known as the \textit{``normalization operator''} or \textit{``jump operator''}. Note that an operator $\hat{A}$ is called a square root of an operator $\hat{B}$ only if it satisfies $\hat{A}^{2}=\hat{B}$. In case that $\hat{B}$ is a positive definite operator, as in the case of $\mathcal{\hat{N}}$, then it possesses a unique positive definite root \cite{matana1,matana2,matana3}. In its defining relation, the operator $\hat{U}^{-1}(t,\,t_{0})$ denotes the inverse of $\hat{U}(t,\,t_{0})$, and~$\hat{U}^{\dagger-1}(t,\,t_{0})$ is the inverse of its adjoint. The inclusion of the operator $\mathcal{\hat{N}}(t,\,t_{0})$ is, in fact, a promotion of the familiar normalization procedure of quantum states to the operatoric level, such that it becomes a local procedure rather than a global one. In other words, instead of having normalization only as an asymptotic procedure, it is now ensured even at the intermediate times.   The unitarity of the above-mentioned solution no longer relies on any assumption with regards to the properties of the space or the Hamiltonian. As will be further explained later, the only formulation of the Schrödinger equation that is consistent with the probabilistic interpretation \cite{Born} is governed by the bounded self-adjoint component. The definition of this component is later provided in Equation~(\ref{decompo}). With $\hat{\mathcal{H}}(t)$ of the general Hamiltonian $\hat{H}(t)$, the resulting equation is solely that of wave mechanics:
\begin{equation}\label{difp}
\frac{d}{dt}\left|\Psi(t)\right\rangle \,=\,-i\hat{\mathcal{H}}(t)\left|\Psi(t)\right\rangle ,\qquad or\qquad\frac{d\hat{\mathcal{P}}(t,\,t_{0})}{dt}\,=\,-i\hat{\mathcal{H}}(t)\,\hat{\mathcal{P}}(t,\,t_{0}).
\end{equation}
\indent  From definition (\ref{definorm}), it is already evident that the introduction of $\mathcal{\hat{N}}$ is not necessary if $\hat{U}$ from (\ref{Udef}) is truly an exact unitary operator. In that case, the adjoint operator coincides with the inverse operator, $\hat{U}^{\dagger}(t,\,t_{0})\,=\,\hat{U}^{-1}(t,\,t_{0})$ and $\hat{U}^{\dagger-1}(t,\,t_{0})\,=\,\hat{U}(t,\,t_{0})$, which leads to $\mathcal{\hat{N}}(t,\,t_{0})=\mathbf{\hat{1}}$. However, as we shall explicitly show, this simplification is generally not valid in various interesting situations; among them is the case of \mbox{gauge~theories}.

This paper is organized as follows: in Section~\ref{trunc}, we explain the reason why the standard solution cannot always provide a reliable unitary description for quantum systems. In Section~\ref{propert}, we provide proof for the validity of the new solution, discuss its properties, and compute the new corresponding perturbative construction. In Section~\ref{exampleuni}, we provide a few practical demonstrations in which the new solution differs from the current mainstream approach. In Section~\ref{conclu}, we summarize the implications of our results on various aspects. In Appendix~\ref{iterative}, we discuss the applicability of the iterative method. In Appendix \ref{dynamn}, the details of the computation of the differential equation for the dynamics of $\hat{\mathcal{N}}$ is provided.

\section{Normalization as a Consequence of Unboundedness}\label{trunc}

In this part, we review the standard argument that is considered proof for the unitarity of $\hat{U}$ in the case of time-dependent Hamiltonian, as mentioned in Equation~(\ref{Udef}). Our intention is to have a closer look at the necessary mathematical conditions involved there, and then discuss the circumstances under which this argument can be invalidated. Afterwards, a new construction is motivated, in which unitarity becomes a robust property and can no longer be broken. And lastly, we characterize, from the mathematical perspective, the different choices of Hamiltonians for which our considerations have become relevant.

\subsection{Why is $\hat{U}$ Not Always Unitary?}\label{notalw}
Our starting point is a quick reminder for the procedure that has led to the result shown in Equation~(\ref{Udef}). The central idea is to solve the differential Equation~(\ref{schrodtime}) by iterations, turning the original differential equation to an integral equation first, and then substituting the result ``inside itself'' repeatedly. This iterative procedure leads to Dyson's series \cite{freema1,freema2},
\begin{equation}\begin{split}\label{uitera}
&\hat{U}(t,\,t_{0})\,=\,\hat{\mathbf{1}}\,-\,i\int_{t_{0}}^{t}dt^{\prime}\:\hat{H}(t^{\prime})\,\hat{U}(t^{\prime},\,t_{0})\\
&\,\;\quad\qquad\,=\,\hat{\mathbf{1}}\,-\,i\int_{t_{0}}^{t}dt^{\prime}\:\hat{H}(t^{\prime})\left[\hat{\mathbf{1}}\,-\,i\int_{t_{0}}^{t^{\prime}}dt^{\prime\prime}\:\hat{H}(t^{\prime\prime})\,\hat{U}(t^{\prime\prime},\,t_{0})\right]\,=\,\sum_{n=0}^{\infty}\hat{u}_{n}(t),
\end{split}\end{equation}
with $\hat{u}_{n}(t)$ denoting a term containing $n$ powers of $\hat{H}(t)$.

At this point the terms of the series above should be interpreted as merely abstract operatoric symbols. The natural way to interpret the operatorial integrations is based on first introducing a complete set as $\mathcal{\hat{O}}(t)\,=\,\int_{\varphi,\psi}\left|\varphi\right\rangle \left\langle \varphi\right|\mathcal{\hat{O}}(t)\left|\psi\right\rangle \left\langle \psi\right|$, where the inner integration includes any $\left|\varphi\right\rangle ,\left|\psi\right\rangle \in\mathcal{D}(\mathcal{\hat{O}})$). Noticeably, in accordance with the solution construction method, it is clear that the original terms of (\ref{uitera}) involve integrations in the iterative form, $\int_{Y}\int_{X}f(x,y)\,dx\,dy$, and not the productive form, $\int_{X\times Y}\,f(x,y)\,dx\,dy$. The applicability of the iterative method is discussed in Appendix~\ref{iterative} but, for now, let us assume the validity of this method. Up to the level of writing explicitly the first two initial non-trivial terms:
\begin{equation}\begin{split}\label{unita}
&\hat{U}(t,\,t_{0})\,=\,\hat{\mathbf{1}}\,-\,i\int_{t_{0}}^{t}dt^{\prime}\:\hat{H}(t^{\prime})\,-\,\int_{t_{0}}^{t}dt^{\prime}\,\hat{H}(t^{\prime})\,\int_{t_{0}}^{t^{\prime}}dt^{\prime\prime}\hat{H}(t^{\prime\prime})\,+\,\ldots
\end{split}\end{equation}

At first sight, the solution (\ref{unita}) seems to satisfy the Schrödinger equation. Indeed, by applying the derivative term by term, it seems that we arrive at
\begin{equation}\label{solvsch}
\frac{d\hat{U}(t,\,t_{0})}{dt}\,=\,\frac{d}{dt}\sum_{n=0}^{\infty}\hat{u}_{n}(t)\,=\,\sum_{n=0}^{\infty}\frac{d\hat{u}_{n}(t)}{dt}\,=\,-i\hat{H}(t)\,\hat{U}(t,\,t_{0}).
\end{equation}

\indent The problem starts when the rules of analysis no longer permits us to conduct the necessary transitions that allow us to claim the validity of (\ref{solvsch}) as the solution. As such is the case when dealing with the regime in which the Leibniz integral rule for interchanging the derivative and integration,
\begin{equation}
\frac{d}{dx}\left(\int_{x_{0}}^{x}dx^{\prime}\,\mathcal{\hat{O}}(x^{\prime})\right)\,=\,\mathcal{\hat{O}}(x),
\end{equation}
is not applicable. Or when interchanging the order of differentiation and summation of the infinite terms,
\begin{equation}
\frac{d}{dt}\sum_{n=0}^{\infty}\hat{u}_{n}(t)\,\rightarrow\,\sum_{n=0}^{\infty}\frac{d}{dt}\hat{u}_{n}(t),
\end{equation}
is generally not permitted \cite{func1}. The last series might be non-differentiable, or the series of derivatives might not converge, or the series of derivatives converges into something other than the derivatives of the series. The convergence properties of the terms and of the series as a whole are dictated by the choice of Hamiltonian, as well as the time interval of the evolution. We now present the familiar argument that currently exists in the literature and is regarded as a proof for the unitarity of $\hat{U}(t,t_{0})$. Assuming that one can replace the integrations of (\ref{uitera}) from the iterative to the productive form,
\begin{equation}\label{uexp1}
\hat{U}(t,\,t_{0})\,=\,\hat{\mathbf{1}}\,-\,i\int_{t_{0}}^{t}dt^{\prime}\:\hat{H}(t^{\prime})\,-\,\int_{t_{0}}^{t}dt^{\prime}\int_{t_{0}}^{t^{\prime}}dt^{\prime\prime}\,\hat{H}(t^{\prime})\,\hat{H}(t^{\prime\prime})\,+\,\ldots
\end{equation}

By further \textls[-25]{assuming $\hat{H}^{\dagger}(t)\,=\,\hat{H}(t)$, along with the validity of the operator simplifications $\left(\int_{t_{0}}^{t}dt^{\prime}\,\hat{H}(t^{\prime})\right)^{\dagger}=\int_{t_{0}}^{t}dt^{\prime}\,\hat{H}(t^{\prime});\;\left(\int_{t_{0}}^{t}dt^{\prime}\int_{t_{0}}^{t^{\prime}}dt^{\prime\prime}\,\hat{H}(t^{\prime})\hat{H}(t^{\prime\prime})\right)^{\dagger}=\int_{t_{0}}^{t}dt^{\prime}\int_{t_{0}}^{t^{\prime}}dt^{\prime\prime}\,\hat{H}(t^{\prime\prime})\hat{H}(t^{\prime})$}, along with
\begin{equation}
(\hat{\mathcal{O}}_{1}+\hat{\mathcal{O}}_{2})^{\dagger}\,=\,\hat{\mathcal{O}}_{1}^{\dagger}+\hat{\mathcal{O}}_{2}^{\dagger},\qquad\qquad(\hat{\mathcal{O}}_{1}\,\hat{\mathcal{O}}_{2})^{\dagger}\,=\,\hat{\mathcal{O}}_{2}\,\hat{\mathcal{O}}_{1}.
\end{equation}

The expansion conjugate to (\ref{uexp1}) takes the form:
\begin{equation}\label{uexp2}
\hat{U}^{\dagger}(t,t_{0})\,=\,\hat{\mathbf{1}}\,+\,i\int_{t_{0}}^{t}dt^{\prime}\:\hat{H}(t^{\prime})\,-\,\int_{t_{0}}^{t}dt^{\prime}\int_{t_{0}}^{t^{\prime}}dt^{\prime\prime}\:\hat{H}(t^{\prime\prime})\,\hat{H}(t^{\prime})\,+\,\ldots
\end{equation} 

Then, by~taking the product of (\ref{uexp1}) and (\ref{uexp2}) the following result is obtained:
\begin{equation}\begin{split}\label{uudag}
&\hat{U}^{\dagger}(t,t_{0})\,\hat{U}(t,t_{0})\,=\,\mathbf{\hat{1}}\,+\,\left(\int_{t_{0}}^{t}dt^{\prime}\:\hat{H}(t^{\prime})\right)^{2}\,-\,\int_{t_{0}}^{t}dt^{\prime}\int_{t_{0}}^{t^{\prime}}dt^{\prime\prime}\:\hat{H}(t^{\prime})\,\hat{H}(t^{\prime\prime})\\
&\qquad\qquad\qquad\quad\;-\,\int_{t_{0}}^{t}dt^{\prime}\int_{t_{0}}^{t^{\prime}}dt^{\prime\prime}\:\hat{H}(t^{\prime\prime})\,\hat{H}(t^{\prime})\,+\,\ldots
\end{split}\end{equation} 

\begin{figure}[H]
\center
  \includegraphics[scale=0.8]{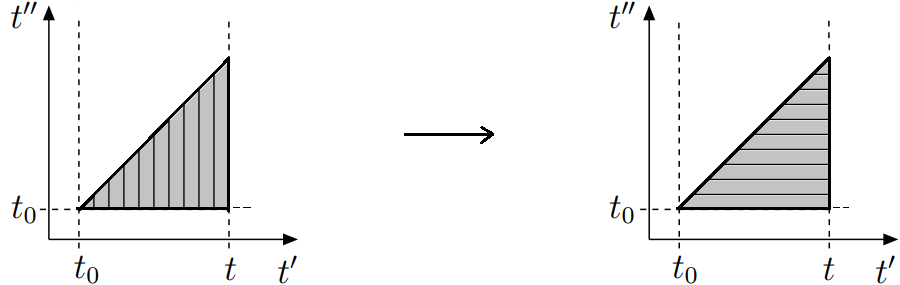}
  \caption{Changing the order of integration via the Fubini theorem. Left: Original integration; this integration was required to be performed on the triangle defined by first integrating the interval $t^{\prime\prime}\in[t_{0},\,t^{\prime}]$, and only then $t^{\prime}\in[t_{0},\,t]$. Right: Integration after applying the Fubini theorem---the first integration is carried out on the interval $t^{\prime\prime}\in[t^{\prime},\,t]$.}\label{fubibi}
\end{figure}

The next step, as shown in Figure~\ref{fubibi}, is to apply an ``operatoric version'' of the Fubini theorem~\cite{Rudin1,Rudin2,Rudin3,Rudin4} and exchange the ordering of the two time integrations in the last term of~(\ref{uudag}),
\begin{equation}\label{unigara}
\int_{t_{0}}^{t}dt^{\prime}\int_{t_{0}}^{t^{\prime}}dt^{\prime\prime}\:\hat{H}(t^{\prime\prime})\,\hat{H}(t^{\prime})\,=\,\int_{t_{0}}^{t}dt^{\prime}\int_{t^{\prime}}^{t}dt^{\prime\prime}\:\hat{H}(t^{\prime})\,\hat{H}(t^{\prime\prime}).
\end{equation}

After introducing the replacement (\ref{unigara}) in the result of (\ref{uudag}), the last two terms can be combined together,
\begin{equation}\label{wrosimp}
\int_{t_{0}}^{t}dt^{\prime}\int_{t_{0}}^{t^{\prime}}dt^{\prime\prime}\:\hat{H}(t^{\prime})\,\hat{H}(t^{\prime\prime})\,+\,\int_{t_{0}}^{t}dt^{\prime}\int_{t^{\prime}}^{t}dt^{\prime\prime}\:\hat{H}(t^{\prime})\,\hat{H}(t^{\prime\prime})\,=\,\left(\int_{t_{0}}^{t}dt^{\prime}\,\hat{H}(t^{\prime})\right)^{2},
\end{equation}
which allegedly seems to lead to the inevitable conclusion that $\hat{U}^{\dagger}(t,\,t_{0})\,=\,\hat{U}^{-1}(t,\,t_{0})$.\\
\vspace{-6pt}
Now, let us look again at the above-mentioned argument but adopt a more cautious approach this time, paying attention to the mathematical conditions involved at each step. Before we proceed, a couple of wise questions to ask ourselves would be the following: 

\textit{Is this argument applicable for the case of an unbounded Hamiltonian? How does the domain of the Hamiltonian, $D(\hat{H})$, affect the argument?}

A detailed discussion about the differences between these two kinds of mathematical constructions can be found in~\cite{Tes1,Tes2}. From there, we learn that confidently answering the question above is necessary in order to determine a legitimate set of simplification operations that can be performed. For clarity, let us summarize the underlying requirements which were highlighted while presenting the unitarity argument above:

$\diamondsuit$ \textbf{Additivity of integrands}: for operators $\mathcal{\hat{O}}_{1}$ and $\mathcal{\hat{O}}_{2}$, 
\begin{equation}\label{lineari}
\int_{t_{0}}^{t_{1}}dt\,\mathcal{\hat{O}}_{1}(t)\,+\,\int_{t_{0}}^{t_{1}}dt\,\mathcal{\hat{O}}_{2}(t)\,=\,\int_{t_{0}}^{t_{1}}dt\,\left(\mathcal{\hat{O}}_{1}(t)+\mathcal{\hat{O}}_{2}(t)\right).
\end{equation}

$\diamondsuit$ \textbf{Additivity of integration intervals}: for an operator $\mathcal{\hat{O}}$,
\begin{equation}\label{additi}
\int_{t_{0}}^{t_{1}}dt\:\mathcal{\hat{O}}(t)\,+\,\int_{t_{1}}^{t_{2}}dt\:\mathcal{\hat{O}}(t)\,=\,\int_{t_{0}}^{t_{2}}dt\:\mathcal{\hat{O}}(t)\,.
\end{equation}

Our ability to perform operation (\ref{lineari}) is mathematically justified only if the domains of the integrated operators on both sides is equivalent throughout the evolution. Contrary to the case of bounded operators, unbounded operators within a given space do not form an algebra or even a complete linear space~\cite{unbounded1,unbounded2}. Each unbounded operator is defined on its own domain, such that if $\mathcal{\hat{O}}_{1}$ and $\mathcal{\hat{O}}_{2}$ are two unbounded operators defined on the domains $\mathcal{D}(\mathcal{\hat{O}}_{1})$ and $\mathcal{D}(\mathcal{\hat{O}}_{2})$, respectively, then the domain of operator $\mathcal{\hat{O}}_{1}+\mathcal{\hat{O}}_{2}$ would be $\mathcal{D}(\mathcal{\hat{O}}_{1})\cap\mathcal{D}(\mathcal{\hat{O}}_{2})$. Note that two operators which act in the same way are to be considered to be different if they are not defined on the same subspace of Hilbert space. It worth mentioning that according to the Hellinger--Toeplitz theorem~\cite{func1}, if a self-adjoint operator is well defined on the entire Hilbert space, it has to be bounded. More importantly, in order to satisfy property (\ref{additi}), the operator $\mathcal{\hat{O}}(t_{1})$ is required to be a bounded operator. It is worth mentioning that one can ``save the additivity property'' by replacing the standard Riemann integral with a modified definition of an integral, but this will not cure the fundamental problem; it will just be hidden inside the modified integral definitions.
A sufficient condition that ensures the validity of the unitarity argument is that the involved operators are satisfying an analogous condition to absolute the convergence for any choice of $\left|\psi\right\rangle \in\mathcal{D}(\hat{H})$,
\begin{equation}\label{orderg2}
\int_{t_{0}}^{t}dt^{\prime}\int_{t_{0}}^{t^{\prime}}dt^{\prime\prime}\:\left\Vert \hat{H}(t^{\prime\prime})\,\hat{H}(t^{\prime})\,\left|\psi\right\rangle \right\Vert _{\mathcal{Y}}^{2}\,<\,\infty\:.
\end{equation}
\indent However, in various interesting problems (such as gauge theories) we often deal with situations in which the condition above is violated. That happens either since the time interval under consideration is improper, $\Delta t\equiv t-t_{0}=\infty$, or due to the integrand containing singularities, as discussed in Section \ref{combi}. In these cases, the resulting integrals (potentially obtained after the regularization procedure) converge only conditionally and demand careful treatment.
\subsubsection*{What About Stone's Theorem?}\label{unitariz}
In the case that the Hamiltonian is time independent, the theorem that allows us to regard (\ref{Udefsimple}) as a unitary operator is given by Stone's theorem~\cite{stone1,stone2}. This theorem provides a one-to-one correspondence between the self-adjoint and unitary operators as stated below.\\

$\bullet$ {\textbf{Stone's theorem:} Let $\hat{H}:\;\mathcal{D}(\hat{H})\rightarrow\mathcal{H}$ be a possibly unbounded self-adjoint operator, then the map $\hat{U}(\Delta t)\,=\,e^{-i\Delta t\,\hat{H}}$ is a strongly continuous, one-parameter unitary operator.}\\
 
However, there are saddle situations in which one can no longer rely on this theorem due to the requirement of being extra cautious when treating the element $\infty$. Below we list a list of three examples:

$\diamondsuit$ \textbf{Infinitely dimensional Hilbert spaces}: As discussed in Section \ref{freepar}, a self-adjoint operator satisfies an additional condition on its domain that an Hermitian operator does not. In the case where $\mathcal{D}(\hat{H}^{\dagger})\,\neq\,\mathcal{D}(\hat{H})$, it can be said that the Hamiltonian is Hermitian but not self-adjoint. Note that this type of space is fundamental to QM since, in order to satisfy the relation, $[\hat{x},\,\hat{p}]\,=\,i\hbar$, either the position $\hat{x}$ or momentum operators $\hat{p}$ must be an unbounded operator. Otherwise, by tracing both sides of the relation, an illogical result is obtained~\cite{Gieres1,Gieres2,Gieres3}. In order to apply Stone's theorem, the property of Hermiticity alone is an insufficient condition, so one should not be surprised that establishing unitarity cannot be~guaranteed.

$\diamondsuit$ \textbf{Singular potentials}: These type of potentials are defined only on a densely defined domain but not everywhere. An example for this choice of Hamiltonian is explicitly discussed in Section \ref{coulomb},
\begin{equation}\label{singulham}
\hat{H}(x)\,=\,\frac{1}{\left\Vert \hat{x}-\hat{x}_{0}\right\Vert }\,,
\end{equation}
which is defined on the domain $\mathcal{D}(\hat{H})\,=\,\left\{ \left\Vert \hat{H}\left|\psi\right\rangle \right\Vert <\infty\right\}.$ It is common to consider this problem within an inadequate space, $\mathcal{H}\,=\,L^{2}(\mathbb{R})$, in which the value of the Hamiltonian on the singular location is replaced by $0$. However, this cannot be performed for a general choice of space, such as indefinite norm spaces. In that case, Stone's theorem cannot imply unitarity in the region outside the domain of the Hamiltonian,
\begin{equation}
\mathcal{E}(\hat{H})\,\equiv\,\mathcal{H}\setminus\mathcal{D}(\hat{H})\,=\,\left\{ \left\Vert \hat{H}\left|\psi\right\rangle \right\Vert =\infty\right\}.
\end{equation}
\indent In other words, unitarity with Hamiltonian (\ref{singulham}) cannot be shown on $\mathcal{E}$. That happens since the state $\left|\psi\right\rangle$, satisfying $\hat{x}\left|\psi\right\rangle \,=\,x_{0}\left|\psi\right\rangle$, does not evolve in a strongly continuous way.

$\diamondsuit$ \textbf{Asymptotic states}: The case in which $t-t_{0}=\infty$, as typically involved in the definition of the $S$-matrix. In such a case, the limit operation is not commutative with the~product,
\begin{equation}
\lim_{t_{1}-t_{0}\rightarrow\infty}\hat{U}(t_{1},t_{0})\,\lim_{t_{1}-t_{0}\rightarrow\infty}\hat{U}^{\dagger}(t_{1},t_{0})\neq\lim_{t_{1}-t_{0}\rightarrow\infty}\hat{U}(t_{1},t_{0})\,\hat{U}^{\dagger}(t_{1},t_{0})\,.
\end{equation}

\subsection{How to Unitarize $\hat{U}$?}\label{unitariz}
The main point of the last section was that unitarity is not a manifest property of $\hat{U}$ for a general choice of space, but rather a property that is implied only under a certain restrictive assumptions. An advancement toward a manifestly unitary solution can be carried out by introducing the replacement:
\begin{equation}
\mathbf{\mathcal{\hat{O}}}\,\longrightarrow\,\sqrt{\mathbf{\mathcal{\hat{O}}}^{\dagger-1}\,\mathbf{\mathcal{\hat{O}}}^{-1}}\,\mathbf{\mathcal{\hat{O}}},
\end{equation}
which leads to the proposed solution in (\ref{pitaron}). This decomposition is analogous to polar decomposition, which always exists and is always unique for a positive definite root choice~\cite{polar1,polar2}. By construction, the solution $\mathcal{\hat{P}}(t,\,t_{0})$ manifestly preserves the exactness of unitarity at all orders and at all times,
\begin{equation}\label{fullunitar}
\mathcal{\hat{P}}^{\dagger}(t,\,t_{0})\,\mathcal{\hat{P}}(t,\,t_{0})\,=\,\hat{U}^{\dagger}(t,\,t_{0})\,\hat{U}^{\dagger-1}(t,\,t_{0})\,\hat{U}^{-1}(t,\,t_{0})\,\hat{U}(t,\,t_{0})\,=\,\boldsymbol{\hat{1}},
\end{equation}
from which it is immediately apparent that $\mathcal{\hat{P}}^{\dagger}(t,\,t_{0})\,=\,\mathcal{\hat{P}}^{-1}(t,\,t_{0})\,$.
The expression for the inverse of $\hat{U}$ can be uniquely determined by the condition $\hat{U}^{-1}(t,\,t_{0})\,\hat{U}(t,\,t_{0})\,=\,\hat{\boldsymbol{1}}$, that leads to
\begin{equation}\label{uinv}
\hat{U}^{-1}(t,t_{0})\,=\,\hat{\mathbf{1}}+i\int_{t_{0}}^{t}dt^{\prime}\:\hat{H}(t^{\prime})-\left(\int_{t_{0}}^{t}dt^{\prime}\,\hat{H}(t^{\prime})\right)^{2}+\int_{t_{0}}^{t}dt^{\prime}\,\hat{H}(t^{\prime})\int_{t_{0}}^{t^{\prime}}dt^{\prime\prime}\,\hat{H}(t^{\prime\prime})\,+\,\ldots
\end{equation}
\indent After insertion back to (\ref{definorm}), along with the Taylor expansion of the square root, the following result is obtained for the case of a bounded self-adjoint Hamiltonian:
\begin{equation}\begin{split}\label{exppnor}
&\mathcal{\hat{N}}(t,\,t_{0})\,=\,\mathbf{\hat{1}}\,-\,\frac{1}{2}\left(\int_{t_{0}}^{t}dt^{\prime}\,\hat{H}(t^{\prime})\right)^{2}\,+\,\frac{1}{2}\int_{t_{0}}^{t}dt^{\prime}\int_{t_{0}}^{t^{\prime}}dt^{\prime\prime}\,\left\{ \hat{H}(t^{\prime}),\,\hat{H}(t^{\prime\prime})\right\}\,+\,\ldots
\end{split}\end{equation}

Then, by using the definition (\ref{pitaron}), we find the expansion:
\begin{equation}\begin{split}\label{pitarexp}
&\mathcal{\hat{P}}(t,\,t_{0})\\
&=\,\mathbf{\hat{1}}\,-\,i\int_{t_{0}}^{t}dt^{\prime}\:\hat{H}(t^{\prime})\,-\,\frac{1}{2}\left(\int_{t_{0}}^{t}dt^{\prime}\:\hat{H}(t^{\prime})\right)^{2}\,-\,\frac{1}{2}\int_{t_{0}}^{t}dt^{\prime}\int_{t_{0}}^{t^{\prime}}dt^{\prime\prime}\:\left[\hat{H}(t^{\prime}),\,\hat{H}(t^{\prime\prime})\right]\,+\,\ldots
\end{split}\end{equation}

More generally, the original terms must be kept in their full glory and the approximation $\sqrt{1+ix-x^{2}}\,\approx\,1+\frac{i}{2}x-\frac{3}{8}x^{2}$ is provides us with
\begin{equation}\begin{split}\label{tun}
&\hat{\mathcal{N}}(t,t_{0})\,=\,\hat{\mathbf{1}}+\frac{i}{2}\int_{t_{0}}^{t}dt^{\prime}\:\hat{H}(t^{\prime})-\frac{i}{2}\left(\int_{t_{0}}^{t}dt^{\prime}\:\hat{H}(t^{\prime})\right)^{\dagger}-\frac{3}{8}\left(\int_{t_{0}}^{t}dt^{\prime}\,\hat{H}(t^{\prime})\right)^{2}\\
&-\frac{3}{8}\left(\int_{t_{0}}^{t}dt^{\prime}\,\hat{H}(t^{\prime})\right)^{\dagger\,2}+\frac{1}{4}\left|\int_{t_{0}}^{t}dt^{\prime}\,\hat{H}(t^{\prime})\right|^{2}+\frac{1}{2}\int_{t_{0}}^{t}dt^{\prime}\hat{H}(t^{\prime})\int_{t_{0}}^{t^{\prime}}dt^{\prime\prime}\hat{H}(t^{\prime\prime})\\
&+\frac{1}{2}\left(\int_{t_{0}}^{t}dt^{\prime}\hat{H}(t^{\prime})\int_{t_{0}}^{t^{\prime}}dt^{\prime\prime}\hat{H}(t^{\prime\prime})\right)^{\dagger}\,+\,\ldots
\end{split}\end{equation}

In the result above the notation $\left|\hat{\boldsymbol{A}}\right|\equiv\sqrt{\hat{\boldsymbol{A}}^{\dagger}\,\hat{\boldsymbol{A}}}$ is introduced (notice the following difference with the definition of the norm: The outcome of $\left|\cdotp\right|$ is another operator, while the operation $\left\Vert \cdotp\right\Vert $ includes an additional tracing operation, leaving us with just a number). Finally, in accordance with (\ref{pitaron}), we arrive at a manifestly unitary result,
\begin{equation}\begin{split}\label{pitarg2}
&\hat{\mathcal{P}}(t,t_{0})\,=\,\hat{\mathbf{1}}-\frac{i}{2}\int_{t_{0}}^{t}dt^{\prime}\:\hat{H}(t^{\prime})-\frac{i}{2}\left(\int_{t_{0}}^{t}dt^{\prime}\:\hat{H}(t^{\prime})\right)^{\dagger}+\frac{1}{8}\left(\int_{t_{0}}^{t}dt^{\prime}\:\hat{H}(t^{\prime})\right)^{2}\\
&-\frac{3}{8}\left(\int_{t_{0}}^{t}dt^{\prime}\:\hat{H}(t^{\prime})\right)^{\dagger\,2}-\frac{1}{4}\left|\int_{t_{0}}^{t}dt^{\prime}\,\hat{H}(t^{\prime})\right|^{2}-\frac{1}{2}\int_{t_{0}}^{t}dt^{\prime}\hat{H}(t^{\prime})\int_{t_{0}}^{t^{\prime}}dt^{\prime\prime}\hat{H}(t^{\prime\prime})\\
&+\frac{1}{2}\left(\int_{t_{0}}^{t}dt^{\prime}\hat{H}(t^{\prime})\int_{t_{0}}^{t^{\prime}}dt^{\prime\prime}\hat{H}(t^{\prime\prime})\right)^{\dagger}\,+\,\ldots
\end{split}\end{equation}

As expected, expansions (\ref{tun}) and (\ref{pitarg2}) are consistent with expressions (\ref{exppnor}) and (\ref{pitarexp}) under the assumption of a bounded self-adjoint Hamiltonian. 

\subsection{Classification of Quantum~Evolution}\label{bounded}
In this part, we discuss the circumstances under which the solution $\hat{U}$ can no longer be considered a complete unitary description. This will be demonstrated by characterizing two types of Hamiltonians and the spaces that typically play a role in the problems of quantum physics.

\subsubsection{The Bounded Self-Adjoint~Evolution}\label{bounham}
In this part, we will provide specific details of the conditions under which our new proposal will not matter 
and the equivalence $\mathcal{\hat{N}}\,=\,\hat{\boldsymbol{1}}$ will be sustained throughout the entire evolution. One of the fundamental postulates in quantum mechanics is that any measurable dynamical quantity is represented by a bounded self-adjoint operator. A crucial property of the Hamiltonians belonging to this category is a spectrum that involves eigenvalues which belong to $\mathbb{R}$. As a simple example for the practical realization of such an operator, it is possible to use a matrix of finite dimensions $\hat{H}(t):L^{2}(\mathbb{R})\rightarrow L^{2}(\mathbb{R})$ with entries given by finite valued and continuous functions. More generally, such a choice of Hamiltonian can be expressed by using a complete orthonormal Hilbert space $\left\{ \left|\phi_{i}\right\rangle \right\}$, 
\begin{equation}
\hat{H}(t)\,=\,\sum_{i=1}^{n}E_{i}(t)\left|\phi_{i}\right\rangle \left\langle \phi_{i}\right|\,+\,\sum_{i,j=1}^{n}f_{ij}(t)\left|\phi_{i}\right\rangle \left\langle \phi_{j}\right|,
\end{equation}
where $E_{i}(t)\in\mathbb{R}$, $f_{ij}(t)=f_{ji}^{*}(t)$ with $\left|E_{i}(t)\right|,\left|f_{ij}(t)\right|<\infty$ for any value of $t$,
\begin{equation}\label{hmilbound}
\hat{H}(t)\,=\,f_{1}\,\hat{\sigma}_{1}+f_{2}\,\hat{\sigma}_{2}+f_{3}\,\hat{\sigma}_{3}\,=\,\left(\begin{array}{cc}
f_{3} & f_{1}-if_{2}\\
f_{1}+if_{2} & -f_{3}
\end{array}\right),
\end{equation}
with $f_{i}=f_{i}(t)\in\mathbb{R}$ and $\left|f_{i}\right|\,<\,\infty$ where $i\in\left\{ 1,2,3\right\}$. In the case where $\Delta t<\infty$, it can be shown that a bounded self-adjoint Hamiltonian implies that
\begin{equation}\label{boundstron}
\left\Vert \left(\hat{U}(t_{2},\,t_{0})\,-\,\hat{U}(t_{1},\,t_{0})\right)\left|\psi(t_{0})\right\rangle \right\Vert _{\mathcal{H}_{2}}\,\leq\,\left\Vert 2\left|\psi(t_{0})\right\rangle \right\Vert _{\mathcal{H}_{1}},
\end{equation}
for any $t_{2},t_{1}\in[t_{0},t]$ and $\left|\psi(t_{0})\right\rangle \in\mathcal{H}_{1}$. The main property that characterizes these Hamiltonians is that the evolution dictated by $\hat{U}$ preserves the norm of the Hilbert space. Thus, for any given $\left|\psi(t_{0})\right\rangle \in\mathcal{H}_{1}$ and action of $\hat{U}$ that generates an isometric transformation, we can observe that
\begin{equation}\label{boundstron2}
\left\Vert \hat{U}(t,\,t_{0})\left|\psi(t_{0})\right\rangle \right\Vert _{\mathcal{H}_{2}}\,=\,\left\Vert \left|\psi(t_{0})\right\rangle \right\Vert _{\mathcal{H}_{1}}.
\end{equation}

\textls[-15]{For time-independent self-adjoint Hamiltonians, the boundness condition ensures the existence of the Taylor series expansion and shows that the construction (\ref{Udefsimple}) is indeed~unitary,}
\begin{equation}\label{unibound}
\left(\exp\left[-i\Delta t\,\hat{H}\right]\right)^{\dagger}=\left(\boldsymbol{\hat{1}}+\sum_{n=1}^{\infty}\frac{(-i\Delta t\,\hat{H})^{n}}{n!}\right)^{\dagger}=\boldsymbol{\hat{1}}+\sum_{n=1}^{\infty}\frac{(i\Delta t\,\hat{H})^{n}}{n!}=\left(\exp\left[i\Delta t\,\hat{H}\right]\right)^{-1}.
\end{equation}
\indent Moreover, even if the Hamiltonian includes time dependence, as long as it is a bounded self-adjoint operator, the argument presented in Section~\ref{trunc} remains valid, and therefore the construction (\ref{Udef}) provides us with a unitary description as well. The reason for that is because the boundness requirement ensures the integrability of the expansion, as guaranteed by the Lebesgue's criterion for integrablility, which states that if $f:\mathbb{R}\rightarrow\mathbb{R}$, then $f$ is Riemann integrable only if $f$ is bounded and the set of discontinuities of $f$ has measure $0$. An alternative way to establish unitarity in this case relies on a simple well-known argument. By multiplying Equation~(\ref{schrodtime}) by $\hat{U}^{\dagger}(t,\,t_{0})$ on the left side, and then multiplying the conjugate equation of motion by $\hat{U}(t,\,t_{0})$ on the right side, we will obtain the equations:
\begin{equation}\begin{split}\label{uandudag}
&\hat{U}^{\dagger}(t,\,t_{0})\,\frac{d\hat{U}(t,\,t_{0})}{dt}\,=\,-i\hat{U}^{\dagger}(t,\,t_{0})\,\hat{H}(t)\,\hat{U}(t,\,t_{0})\,,\\
&\frac{d\hat{U}^{\dagger}(t,\,t_{0})}{dt}\,\hat{U}(t,\,t_{0})\,=\,i\hat{U}^{\dagger}(t,\,t_{0})\,\hat{H}(t)\,\hat{U}(t,\,t_{0})\,.
\end{split}\end{equation}

Adding the two equations in (\ref{uandudag}), one can make use of the Leibniz product rule,
\begin{equation}\label{lieb}
\hat{U}^{\dagger}(t,\,t_{0})\frac{d\hat{U}(t,\,t_{0})}{dt}\,+\,\frac{d\hat{U}^{\dagger}(t,\,t_{0})}{dt}\hat{U}(t,\,t_{0})\,=\,\frac{d}{dt}\left(\hat{U}^{\dagger}(t,\,t_{0})\,\hat{U}(t,\,t_{0})\right)\,=\,0.
\end{equation}

Provided that the initial conditions holds, $\hat{U}^{\dagger}(t_{0},\,t_{0})\,\hat{U}(t_{0},\,t_{0})\,=\,\hat{\boldsymbol{1}}$, this is sufficient to ensure that unitarity is also preserved at all later times,
\begin{equation}
\hat{U}^{\dagger}(t,\,t_{0})\,\hat{U}(t,\,t_{0})\,=\,\boldsymbol{\hat{1}}.
\end{equation}

However, it is important to realize that the above-mentioned argument is crucially based on the operatorical equivalence $\hat{H}^{\dagger}(t)=\hat{H}(t)$ for any $t$, which implies the equivalence of domains,
\begin{equation}
\mathcal{D}(\hat{H}^{\dagger}(t))=\mathcal{D}(\hat{H}(t)).
\end{equation}

An additional property that can be shown for any intermediate time $t_{1}\in[t_{2},\,t_{0}]$ is
\begin{equation}\label{mrko}
\hat{U}(t_{2},\,t_{1})\,\hat{U}(t_{1},\,t_{0})\,=\,\hat{U}(t_{2},\,t_{0}).
\end{equation}

The property above, known as the Markovian property~\cite{Markov1,Markov2,Markov3}, is a characteristic of various stochastic processes. Its intuitive meaning is that the evolution associated with the operator $\hat{U}$ has ``no memory'', or that the evolution from time $t_{1}$ to time $t_{2}$ depends only on the state of the system at time $t_{1}$ and not on the preceding part of the evolution. The generalization of the above is given by the Trotter formula~\cite{Trotter},
\begin{equation}\label{trott}
\lim_{n\rightarrow\infty}\hat{U}(t_{n},\,t_{n-1})\,\cdots\,\hat{U}(t_{2},\,t_{1})\,\hat{U}(t_{1},\,t_{0})\,=\,\lim_{n\rightarrow\infty}\hat{U}(t_{n},\,t_{0}).
\end{equation}

\subsubsection{The Unbounded Evolution}\label{unbounham}
The capacity of the bounded Hamiltonian to describe the quantum phenomena is limited only for systems that evolve in a gradual (functional) way. However, there are a plethora of quantum systems which the described by an unbounded Hamiltonian. As demonstrated in Section \ref{combi}, it can be seen that in these cases the transformation generated by $\hat{U}$ changes the norm of the evolved states,
\begin{equation}\label{chnnorm}
\mathcal{Z}(t,\,t_{0})\,\equiv\,\frac{\left\Vert \hat{U}(t,\,t_{0})\left|\psi(t_{0})\right\rangle \right\Vert _{\mathcal{D}(\hat{H})}}{\left\Vert \left|\psi(t_{0})\right\rangle \right\Vert _{\mathcal{Y}}}\,\neq\,1\,.
\end{equation}
\indent The factor $\mathcal{Z}(t,\,t_{0})$ is typically known as the WF normalization, the LSZ factor, or the field strength. The preservation of the norm is obtained after including the compensation via the action of $\hat{\mathcal{N}}$:
\begin{equation}
\left\Vert \mathcal{\hat{P}}(t,\,t_{0})\left|\psi(t_{0})\right\rangle \right\Vert _{\mathcal{D}(\hat{H})}\,=\,\left\Vert \left|\psi(t_{0})\right\rangle \right\Vert _{\mathcal{Y}}.
\end{equation}

A simple example is the Dirac comb potential (see example in Section \ref{combi}), which involves periodic kicks~\cite{Flugge1,Flugge2}:
\begin{equation}\label{dircom}
H(t)\,=\,\sum_{n=-\infty}^{\infty}\delta(t-nT),
\end{equation}
with $T\in\mathbb{R}$. In order to gain an impression of the subtleties that arise when handling this case we can simply take $H(t)\,=\,\delta(t-T)$. Due to (\ref{undefi}), for such a choice the differentiation and integration are not commutative operations,
\begin{equation}
\frac{d}{dt}\left(\int_{t_{0}}^{t}dt^{\prime}\,\hat{H}(t^{\prime})\,\int_{t_{0}}^{t^{\prime}}dt^{\prime\prime}\hat{H}(t^{\prime\prime})\right)\,\nrightarrow\,\hat{H}(t)\,\int_{t_{0}}^{t}dt^{\prime}\hat{H}(t^{\prime}),
\end{equation}
where the notation of a broken arrow $\nrightarrow$ signifies here an invalid transition. It is also clear that, for this choice, the validity of the Fubini theorem (\ref{unigara}) for the evolution intervals which contain the singular times $T$ has no mathematical justification to be used, as explicitly
\begin{equation}
\int_{t_{0}}^{t}dt^{\prime}\,\delta(t^{\prime}-T)\,\int_{t_{0}}^{t^{\prime}}dt^{\prime\prime}\delta(t^{\prime\prime}-T)\,\nrightarrow\,\int_{t_{0}}^{t}dt^{\prime}\,\delta(t^{\prime}-T)\int_{t^{\prime}}^{t}dt^{\prime\prime}\delta(t^{\prime\prime}-T)\,.
 \end{equation}
\indent In addition, the additivity property (\ref{additi}) holds only as long as $t^{\prime}\neq T$, but it fails when reaching the singular moment $t^{\prime}=T$ as required by the unitarity argument,
\begin{equation}
\int_{t_{0}}^{t^{\prime}=T}dt^{\prime}\,\delta(t^{\prime}-T)\,+\,\int_{t^{\prime}=T}^{t}dt^{\prime}\,\delta(t^{\prime}-T)\,\nrightarrow\,\int_{t_{0}}^{t}dt^{\prime}\,\delta(t^{\prime}-T).
  \end{equation}
\indent Obviously, without these properties, the unitarity argument presented in Section \ref{notalw} cannot be established. At this point, one might be tempted to replace the delta function with its corresponding smeared version in order to avoid these problems. However, this means that our approach is fundamentally incapable of handling singular objects as they really are. As long as $T\notin[t_{0},t]$ is outside of the evolution interval, unitarity is trivially~preserved:
\begin{equation}
\left.U^{\dagger}(t,\,t_{0})\,U(t,\,t_{0})\right|_{T\notin[t_{0},t]}\,=\,1.
\end{equation}
\indent However, for evolution intervals such that $t_{1}\in(\ensuremath{t_{0}},\,t)$, this is no longer the case, as can be observed from:
\begin{equation}
\left.U^{\dagger}(t,\,t_{0})\,U(t,\,t_{0})\right|_{T\in[t_{0},t]}\,\neq\,1,
\end{equation}
\textls[-35]{which is the reason behind (\ref{chnnorm}). Note that contrary to the bounded case, for the unbounded Hamiltonians, there exists at least one value of $t_{1}\in(t,\,t_{0})$ such that $U(t_{2},\,t_{1})\,U(t_{1},\,t_{0})\,\neq\,U(t_{2},\,t_{0}).$}

\section{The Properties of the Pitaron}\label{propert}

In this part, we would like to discuss more about the characteristics of the new solution with regards to the initial conditions and its ability to solve the Schrödinger equation.
\subsection{Satisfying the Initial Conditions}
An immediate observation is that the satisfaction of the initial conditions by $\hat{U}$ implies that they are also satisfied by $\hat{\mathcal{P}}$. Generally speaking, bounded Hamiltonians allow us to satisfy the initial conditions trivially as, in that case, both $\hat{U}(t_{0},\,t_{0})=\hat{\mathbf{1}}$ and $\hat{\mathcal{N}}(t_{0},\,t_{0})=\hat{\mathbf{1}}$ by continuity. Tackling the case of unbounded Hamiltonians is slightly less straightforward, and for $\hat{U}$, one has to make a fair necessary assumption that the initial time does not coincide with the places in which the Hamiltonian becomes singular. For $\mathcal{\hat{P}}$, the necessary condition for ensuring the initial condition becomes
\begin{equation}
\lim_{t\rightarrow t_{0}}\hat{U}(t,\,t_{0})\,=\,\lim_{t\rightarrow t_{0}}\hat{U}^{\dagger}(t,\,t_{0})\quad\longrightarrow\quad\lim_{t\rightarrow t_{0}}\mathcal{\hat{N}}(t,\,t_{0})\,=\,\lim_{t\rightarrow t_{0}}\hat{U}^{-1}(t,\,t_{0}),
\end{equation}
which implies the relation
\begin{equation}
\lim_{t\rightarrow t_{0}}\mathcal{\hat{P}}(t,\,t_{0})=\lim_{t\rightarrow t_{0}}\hat{U}^{-1}(t,\,t_{0})\,\hat{U}(t,\,t_{0})\,=\,\hat{\mathbf{1}}.
\end{equation}
\subsection{Solving the Schrödinger~Equation}\label{tevo}
In this part, we provide mathematical proof for our claim that the solution provided in (\ref{pitaron}) solves the original Schrödinger Equation~(\ref{schrodtime}). Let us start with performing the calculation under the assumption of an Hermitian Hamiltonian. After applying the product rule, the time derivative of $\mathcal{\hat{P}}$ can be expressed as
\begin{equation}\label{derivap}
\frac{d\mathcal{\hat{P}}(t,\,t_{0})}{dt}\,=\,\mathcal{\hat{N}}(t,\,t_{0})\frac{d\hat{U}(t,\,t_{0})}{dt}\,+\,\frac{d\mathcal{\hat{N}}(t,\,t_{0})}{dt}\hat{U}(t,\,t_{0}).
\end{equation}

It is worth mentioning that the differentiation and conjugation are not commutative operations, $\frac{d}{dt}\mathcal{\hat{P}}^{\dagger}(t,\,t_{0})\,\neq\,\left(\frac{d}{dt}\mathcal{\hat{P}}(t,\,t_{0})\right)^{\dagger}$, with equivalence only when using a self-adjoint Hamiltonian. By taking the time derivative of Equations~(\ref{unita}) and (\ref{exppnor}), one arrives at the following set of equations:
\begin{equation}\begin{split}\label{boundedeqs}
&\frac{d\hat{U}(t,\,t_{0})}{dt}\,=\,-i\hat{H}(t)\,\hat{U}(t,\,t_{0}),\qquad\frac{d\mathcal{\hat{N}}(t,\,t_{0})}{dt}\,=\,-i\left[\hat{H}(t),\,\mathcal{\hat{N}}(t,\,t_{0})\right].
\end{split}\end{equation}

 After plugging Equation~(\ref{boundedeqs}) inside relation (\ref{derivap}), we note that the Schrödinger equation is indeed satisfied:
\begin{equation}\begin{split}
&\frac{d\mathcal{\hat{P}}(t,\,t_{0})}{dt}\,=\,\mathcal{\hat{N}}(t,\,t_{0})\left(-i\hat{H}(t)\hat{U}(t,\,t_{0})\right)\\
&\qquad\qquad\;+\left(-i\left[\hat{H}(t),\,\mathcal{\hat{N}}(t,\,t_{0})\right]\right)\hat{U}(t,\,t_{0})\,=\,-i\hat{H}(t)\mathcal{\hat{P}}(t,\,t_{0}),
\end{split}\end{equation}
\textls[-10]{where the simplification $\mathcal{\hat{N}}(t,\,t_{0})\hat{H}(t)\,+\,\left[\hat{H}(t),\,\mathcal{\hat{N}}(t,\,t_{0})\right]\,=\,\hat{H}(t)\mathcal{\hat{N}}(t,\,t_{0})$ is used. One recognizes that the resulting equation for $\mathcal{\hat{N}}(t,\,t_{0})$ is, in fact, the quantum Liouville equation \cite{Bogolubov}, known alternatively as the von Neuman equation~\cite{Neumann}}. This equation ensures the probability conservation in our phase-space, and more commonly written in the equivalent~form
\begin{equation}\begin{split}
&\frac{d\mathcal{\hat{N}}^{2}(t,\,t_{0})}{dt}\,=\,-i\mathcal{\hat{N}}(t,\,t_{0})\left[\hat{H}(t),\,\mathcal{\hat{N}}(t,\,t_{0})\right]-i\left[\hat{H}(t),\,\mathcal{\hat{N}}(t,\,t_{0})\right]\mathcal{\hat{N}}(t,\,t_{0})\\
&\qquad\quad\quad\;\;\:=\,-i\left[\hat{H}(t),\,\mathcal{\hat{N}}^{2}(t,\,t_{0})\right].
\end{split}\end{equation}
\indent The dynamics of the above Liouville equation can be observed as trivial: the value of $\mathcal{\hat{N}}(t_{0},\,t_{0})=\hat{\boldsymbol{1}}$ remains unchanged throughout the entire evolution. However, this will no longer be justified in the case where the Hamiltonian is unbounded, or alternatively, non-Hermitian \cite{Moiseyev1,Moiseyev2,Moiseyev3,Moiseyev4,Moiseyev5,Moiseyev6,Moiseyev7}. The Hamiltonians which are under consideration here can be decomposed as a sum of a self-adjoint component and a singular component. Note that in order for operators $\hat{\mathcal{H}}(t)$ and $\mathcal{\hat{J}}(t)$ to represent an eligible decomposition of a diagonalizible Hamiltonian, they need to share a common set of eigenvectors. In that case, they are simultaneously diagonalizable, which implies the relations $\left[\hat{\mathcal{H}}(t),\,\mathcal{\hat{J}}(t)\right]\,=\,0$. The singular component is non-vanishing on a measure-0 set and exists as a generalized object (see example Section \ref{coulomb}),
\begin{equation}\label{decompo}
\hat{H}(t)\,=\,\hat{\mathcal{H}}(t)-i\mathcal{\hat{J}}(t).
\end{equation}

The corresponding evolution equations (see Appendix \ref{dynamn}) are:
\begin{equation}\begin{split}\label{twoeqs}
&\frac{d\hat{U}(t,\,t_{0})}{dt}\,=\,-i\left(\hat{\mathcal{H}}(t)-i\mathcal{\hat{J}}(t)\right)\,\hat{U}(t,\,t_{0})\,,\\
&\frac{d\mathcal{\hat{N}}(t,\,t_{0})}{dt}\,=\,-i\left[\hat{\mathcal{H}}(t),\,\mathcal{\hat{N}}(t,\,t_{0})\right]\,+\,\mathcal{\hat{N}}(t,\,t_{0})\,\mathcal{\hat{J}}(t).
\end{split}\end{equation}
\indent It is straightforward to verify that Equation~(\ref{twoeqs}) above indeed solves the time-dependent Schrödinger Equation~(\ref{schrodtime}) by inserting it into (\ref{derivap}),
\begin{equation}\begin{split}\label{dersch}
&\frac{d\mathcal{\hat{P}}(t,\,t_{0})}{dt}\,=\,\mathcal{\hat{N}}(t,\,t_{0})\left(-i\left(\hat{\mathcal{H}}(t)-i\mathcal{\hat{J}}(t)\right)\,\hat{U}(t,\,t_{0})\right)\\
&\qquad\qquad\;\,+\,\left(-i\left[\hat{\mathcal{H}}(t),\,\mathcal{\hat{N}}(t,\,t_{0})\right]\,+\,\mathcal{\hat{N}}(t,\,t_{0})\,\mathcal{\hat{J}}(t)\right)\hat{U}(t,\,t_{0})\,=\,-i\hat{\mathcal{H}}(t)\,\mathcal{\hat{P}}(t,\,t_{0})\,.
\end{split}\end{equation}
\indent Unsurprisingly, the~RHS of the above equation no longer depends on the anti-self-adjoint component, $i\mathcal{\hat{J}}(t)$. This should have been expected as, unlike the case of $\hat{U}$, the differential equation for $\mathcal{\hat{P}}$ does not predict the decay of the probabilistic wave function with time, even in the presence of a complex/non-Hermitian Hamiltonian. In a case where the component $\hat{\mathcal{H}}(t)$ is bounded and evaluated on a proper time interval, the time evolution operator based on (\ref{dersch}) can be constructed analogously to (\ref{Udef}),
\begin{equation}
\hat{\mathcal{P}}(t,\,t_{0})\,=\,\hat{T}\exp\left[-i\int_{t_{0}}^{t}dt^{\prime}\,\hat{\mathcal{H}}(t^{\prime})\right]\,.
\end{equation}
\indent Note that according to the traditional approach, in~situations in which the Hamiltonian is Hermitian but not self-adjoint, the~replacement $\hat{U}\rightarrow\hat{\mathcal{N}}\hat{U}$ should not affect the corresponding differential equation; however, in reality, we can see that it does.

 \section{Examples}\label{exampleuni}

\begin{figure}[H]
\center
  \includegraphics[scale=0.6]{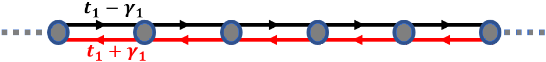}
  \caption{The non-Hermitian skin effect (NHSE) Hamiltonian. At each site the particle can either transit to its left or right nearby site with an unequal probabilities.\label{skin}}
\end{figure}

In this part, we demonstrate with simple examples how our proposed solution affects the description of quantum systems practically. Even without studying any particular example, one can observe that when truncating the perturbative expansion at the Born approximation unitarity becomes broken. A necessity of normalizing such an expansion in order to maintain unitarity is apparent. In fact, similar consideration will apply whenever truncating the expansion for $\hat{U}(t,\,t_{0})$ at any level involving odd powers of $\hat{H}$. Our intention here is to demonstrate situations where unitarity is absent even when dealing with an even order of powers of $\hat{H}$. Alternatively, a straightforward way to obtain a non-trivial modification is when dealing with a Hamiltonian that is explicitly non-Hermitian, as in (\ref{decompo}). In that case, one finds:
\begin{equation}\label{nonher}
\hat{U}(t,t_{0})\,=\,\exp\left(-i(\hat{\mathcal{H}}-i\hat{\mathcal{J}})\Delta t\right),\qquad\hat{\mathcal{N}}(t,t_{0})\,=\,\exp\left(\hat{\mathcal{J}}\Delta t\right),
\end{equation}
and $\hat{\mathcal{P}}(t,t_{0})=\exp\left(\hat{\mathcal{J}}\Delta t\right)\exp\left(-i(\hat{\mathcal{H}}-i\hat{\mathcal{J}})\Delta t\right)=\exp\left(-i\hat{\mathcal{H}}\Delta t\right)$. As an explicit demonstration model, one can use the non-Hermitian skin effect (NHSE), where a macroscopic number of particles are initially accumulated at a boundary of a lattice of size $l$. These particles are allowed to propagate on the lattice with unequal probabilities (see Figure \ref{skin}) according to the Hamiltonian
\begin{equation}
\hat{H}\,=\,\sum_{i=1}^{l}Ea_{i}^{\dagger}a_{i}+\sum_{i=1}^{l-1}\left[(t_{i}-\gamma_{i})a_{i}^{\dagger}a_{i+1}+(t_{i}+\gamma_{i})a_{i+1}^{\dagger}a_{i}\right].
\end{equation}

In this definition $a_{i}$ and $a_{i}^{\dagger}$ are the creation and annihilation operators of a particle in the $i$th unit cell. Obviously, the interaction term breaks Hermiticity due to the finite size of the lattice and the asymmetric nature of the probabilities for a left/right transition ($t_{i}+\gamma_{i}$ vs $t_{i}-\gamma_{i}$), and one observes that $\hat{H}^{\dagger}\,\neq\,\hat{H}$. The current approach will regard the time evolution of this system as a dissipation system, such that $\Delta t\rightarrow\infty$ implies $\left|\psi(t)\right\rangle \rightarrow0$. On the other hand, in the proposed formalism the probability after the each step is getting re-distributed such that unitarity is always preserved manifestly. Thus, our suggested approach allows to bring the usual understanding of Born's probabilistic interpretation \cite{Born} for these systems. In the additional examples discussed in this section, we show that obtaining a non-trivial effect out of $\hat{\mathcal{N}}$ is, in fact, more common than one can originally expect.
\subsection{The Free~Particle}\label{freepar}
Generally, a~particle subject to a time-independent Hamiltonian is given as a sum of its kinetic and potential energy,
\begin{equation}
\hat{H}(\hat{x},\hat{p})\,=\,\frac{\hat{p}^{2}}{2m}\,+\,\hat{V}(\hat{x},\hat{p}).
 \end{equation}
\indent A simple choice is to examine the case of $\hat{V}(\hat{x},\hat{p})=0$, with~a setup such that $x\in[0,1]$. The corresponding Hilbert space is spanned via the complete set of momentum eigenstates such that their corresponding eigenvalues belong to the extended real numbers set $\mathbb{\overline{R}}$,
\begin{equation}
\mathbb{\overline{R}}\,=\,\mathbb{R}\cup\{-\infty,\,\infty\}.
\end{equation}

The resulting differential equation for a particle with mass $m$ is obtained via $\hat{p}\,=\,i\hbar\frac{d}{dx}$,
\begin{equation}
i\hbar\frac{d}{dt}\left|\psi(x,t)\right\rangle \,=\,-\frac{\hbar^{2}}{2m}\frac{d^{2}}{dx^{2}}\left|\psi(x,t)\right\rangle. 
 \end{equation}
\indent The kinetic part can be resolved by introducing $\left|\psi(x,t)\right\rangle \,=\,\exp\left[-i\Delta t\,\frac{\hat{p}^{2}}{2m}\right]\left|\psi(x,t_{0})\right\rangle $, from which we infer the time evolution operator
\begin{equation}\label{freeU}
\hat{U}(t,\,t_{0})\,=\,\exp\left[-i\Delta t\,\frac{\hat{p}^{2}}{2m}\right]\,.
\end{equation}
\indent Our point is that the above operator is Hermitian but not self-adjoint, which is insufficient to generate a unitary transformation. Let us recall that an operator $\mathcal{\hat{O}}$ on the Hilbert space $\mathcal{H}$ is called self-adjoint if it is operatorically identical to its adjoint, $\mathcal{\hat{O}}^{\dagger}=\mathcal{\hat{O}}$. The equivalence between two operators implies the two conditions:
\begin{equation}\label{hermi}
\left\langle \varphi\right|\mathcal{\hat{O}}^{\dagger}\left|\psi\right\rangle =\left\langle \varphi\right|\mathcal{\hat{O}}\left|\psi\right\rangle \qquad\qquad\left|\varphi\right\rangle ,\left|\psi\right\rangle \in\mathcal{H},
\end{equation}
and additionally the domains are identical, 
\begin{equation}
\mathcal{D}(\mathcal{\hat{O}}^{\dagger})=\mathcal{D}(\mathcal{\hat{O}}). 
\end{equation}

The condition above holds automatically when $\mathcal{\hat{O}}$ is a finite-dimensional operator since, in that case, $\mathcal{D}(\mathcal{\hat{O}})=\mathcal{H}$ for every non-singular linear operator within a finite-dimensional space. However, in the infinite-dimensional case, the domain $\mathcal{D}(\mathcal{\hat{O}}^{\dagger})$ may be a larger subspace than $\mathcal{D}(\mathcal{\hat{O}})$, such that $\mathcal{D}(\mathcal{\hat{O}})\subseteq\mathcal{D}(\mathcal{\hat{O}}^{\dagger})$. In that case, where only the condition (\ref{hermi}) is satisfied, the operator $\mathcal{\hat{O}}$ is called Hermitian. Thus, any self-adjoint operator is Hermitian, but a Hermitian operator is not necessarily self-adjoint.
Noticeably, for~a bounded self-adjoint Hamiltonian, the spectrum is real and the eigenvectors associated to different eigenvalues are mutually orthogonal. Moreover, all the eigenvectors have a finite norm and, together, they form a complete Hilbert space~\cite{Hilbert1,Hilbert2}. These properties do not hold for operators which are only Hermitian~\cite{Berezanski}. 

\indent As~demonstrated in~\cite{Gieres1,Gieres2,Gieres3}, for the momentum operator $\hat{p}$ of a particle living on a compact domain $L^{2}[0,1]$, the~equivalence $\left\langle \varphi\right|\hat{p}^{\dagger}\left|\psi\right\rangle =\left\langle \varphi\right|\hat{p}\left|\psi\right\rangle$ holds for any $\left|\varphi\right\rangle ,\left|\psi\right\rangle \in\mathcal{D}(\hat{p})$. This is because for all $\left|\psi(x)\right\rangle \in\mathcal{D}(\hat{p})$, the following applies:
\begin{equation}\begin{split}
&\int_{0}^{1}dx\,\left\langle \varphi(x,t)\right|\hat{p}^{\dagger}\left|\psi(x,t)\right\rangle -\int_{0}^{1}dx\,\left\langle \varphi(x,t)\right|\hat{p}\left|\psi(x,t)\right\rangle \\
&=\,\frac{\hbar}{i}\left(\left\langle \varphi(x=1,t))\left|\psi(x=1,t))\right.\right\rangle -\left\langle \varphi(x=0,t))\left|\psi(x=0,t)\right.\right\rangle \right)\,=\,0\;.
\end{split}\end{equation}

However, since the boundary conditions satisfied by $\left|\psi(x)\right\rangle \in\mathcal{D}(\hat{p})$ are already sufficient for annihilating the surface term,
\begin{equation}
\mathcal{D}(\hat{p})\subset\mathcal{D}(\hat{p}^{\dagger}).
\end{equation}

Thus, the operator $\hat{p}$ is Hermitian but not self-adjoint~\cite{Uwe1,Uwe2}.
\begin{equation}
\hat{p}^{\dagger}\,\neq\,\hat{p}\,.
\end{equation}

The same conclusion remains correct also at the level of the Hamiltonian,
\begin{equation}
\hat{H}^{\dagger}\,=\,\left(\frac{\hat{p}^{2}}{2m}\right)^{\dagger}\,\neq\,\frac{\hat{p}^{2}}{2m}\,=\,\hat{H}.
\end{equation}

\textls[-15]{Therefore, the corresponding time evolution operator (\ref{freeU}) cannot be regarded as~unitary,}
\begin{equation}
\exp^{-1}\left[-i\Delta t\,\frac{\hat{p}^{2}}{2m}\right]\,\neq\,\exp^{\dagger}\left[-i\Delta t\,\frac{\hat{p}^{2}}{2m}\right],
\end{equation}
we arrive at the conclusion
\begin{equation}
\mathcal{\hat{N}}(t,\,t_{0})\,=\,\sqrt{\exp^{\dagger-1}\left[-i\Delta t\,\frac{\hat{p}^{2}}{2m}\right]\,\exp^{-1}\left[-i\Delta t\,\frac{\hat{p}^{2}}{2m}\right]}\,\neq\,\hat{\boldsymbol{1}}\,.
\end{equation}

An alternative way to understand the necessity of $\mathcal{\hat{N}}$ is by realizing that dealing with a bounded space means that there exist a state $\left|p_{max}\right\rangle $ for which
\begin{equation}
\hat{p}\left|p_{max}\right\rangle \,=\,p_{max}\left|p_{max}\right\rangle ,\qquad\qquad\int_{-p_{max}}^{p_{max}}d^{3}p\,\left|p\right\rangle \left\langle p\right|\,=\,\hat{\boldsymbol{1}},
\end{equation}
with $0<p_{max}<\infty$. On the right appears the completeness relation for a vector space with upper eigenvalue of $p_{max}$. On an unbounded space the spectrum of the momentum eigenvalues is the extended real set, and the condition above applies with $p_{max}=\infty$. In terms of eigenstates, the complete momentum space is spanned via the collection of all the states corresponding to the above eigenvalues
\begin{equation}
\left\{ \left|p\right\rangle \right\} \,=\,\left\{ \left|p_{\mathbb{R}}\right\rangle \right\} \cup\left\{ \left|p_{-\infty}\right\rangle ,\,\left|p_{+\infty}\right\rangle \right\} .
\end{equation}

These are defined such that $\hat{p}\left|p_{\mathbb{R}}\right\rangle \,=\,p_{\mathbb{R}}\left|p_{\mathbb{R}}\right\rangle $ with $|p_{\mathbb{R}}|\in\mathbb{R}$, in addition to the boundary momentum states, $\hat{p}\left|p_{\pm\infty}\right\rangle \,=\,\pm\infty\left|p_{\pm\infty}\right\rangle$. The operator $\mathcal{\hat{N}}$ acts trivially on the non-asymptotic momenta states during the non-asymptotic times,
\begin{equation}
\mathcal{\hat{N}}(|\Delta t|<\infty)\left|p_{\mathbb{R}}\right\rangle \,=\,\hat{\boldsymbol{1}}\left|p_{\mathbb{R}}\right\rangle .
\end{equation}

The operator $\mathcal{\hat{N}}$ becomes non-trivial in two cases that involve the element $\infty$:\\
$\diamondsuit$ Asymptotic times: $\:\mathcal{\hat{N}}(|\Delta t|=\infty)\left|p_{\mathbb{R}}\right\rangle \,\neq\,\hat{\boldsymbol{1}}\left|p_{\mathbb{R}}\right\rangle,$\\
$\diamondsuit$ Asymptotic momenta: $\:\mathcal{\hat{N}}(|\Delta t|\leq\infty)\left|p_{\pm\infty}\right\rangle \,\neq\,\hat{\boldsymbol{1}}\left|p_{\pm\infty}\right\rangle$.

\subsection{The Coulomb~Potential}\label{coulomb}
The inter-atomic and molecular potentials are known empirically to be described by the inverse positive powers of the Euclidean norm and therefore cannot be defined everywhere~\cite{Moretti,Solvable}. Well-known examples are the Coulomb, Yukawa, van der Waals, Lennard--Jones, and London potentials~\cite{Flugge1,Flugge2}. Here our interest is to provide a new perspective for analyzing the Coulomb force. Let us start with an explanation of the current approach to dealing with this case, then discuss its drawback and provide an alternative approach. This potential is expressed via
\begin{equation}
V(\hat{r})\,=\,-\,\frac{\kappa}{\left\Vert \hat{r}\right\Vert },\qquad\kappa>0\,,
\end{equation}
and full Hamiltonian, including the kinetic term, is given by
\begin{equation}
H(\hat{r},\hat{p})\,=\,T(\hat{p})+V(\hat{r})\,=\,\frac{\hat{p}^{2}}{2m}-\frac{\kappa}{\left\Vert \hat{r}\right\Vert }\,.
\end{equation}
\indent For simplicity, in order to focus on the part that matters, let us assume that the kinetic term is sufficiently well behaved and that the limit $m\rightarrow\infty$ is taken. In the current approach, an additional assumption that is unjustifiably introduced is that the problem can be adequately studied on the space $L^{2}(\mathbb{R})$. Under such an assumption, contributions of measure-0 has no importance, and the singularity at the origin can be disregarded. Thus, it is sufficient to specify the potential only on a densely defined part of the space,
\begin{equation}\label{nomeas}
V(r)\,=\,\begin{cases}
\begin{array}{c}
-\,\frac{\kappa}{r}\\
0
\end{array} & \begin{array}{c}
r\neq0\\
r=0
\end{array}\end{cases}\,.
\end{equation}
\indent As expected, the corresponding time evolution operator, Equation~(\ref{Udefsimple}), based on representation (\ref{nomeas}) is unitary everywhere. However, expression (\ref{nomeas}) is an over-simplification of the original singular potential and is incompatible with the classically correspondent case in which the source is to be represented by a distribution. Suppose that we no longer assume that we are dealing with $L^{2}(\mathbb{R})$, then our previous assumption that measure-0 contributions can be ignored is no longer justified. 
The first thing we have to do is provide a meaning to the singularity in terms of familiar objects. This can be carried out by generalizing the framework, $\mathbb{R}\rightarrow\mathbb{C}$, such that $V(r)\rightarrow V(r,\epsilon)=\frac{1}{r-i\epsilon}$. Eventually, we can return to the real line using the vanishing imaginary component, $\epsilon\rightarrow0$. A valuable relation in our case is the Sokhotski theorem,
\begin{equation}
V(r)=\lim_{\epsilon\rightarrow0}V(r,\epsilon)=\lim_{\epsilon\rightarrow0}\frac{1}{r-i\epsilon}\,=\,P.v.\left(\frac{1}{r}\right)+i\pi\delta(r).
\end{equation}

Thus, in our approach, the consistent representation of the Coulomb force is
\begin{equation}
V(r)\,=\,\begin{cases}
\begin{array}{c}
-\,\frac{\kappa}{r}\\
-\,i\pi\kappa\,\delta(r)
\end{array} & \begin{array}{c}
r\neq0\\
r=0
\end{array}\end{cases}.
\end{equation}

The corresponding expression for the time evolution operator is
\begin{equation}
U(r,\,\Delta t)=\begin{cases}
\begin{array}{c}
\exp\left[-\frac{i\kappa}{r}\Delta t\right]\\
\exp\left[\kappa\pi\delta(r)\,\Delta t\right]
\end{array} & \begin{array}{c}
r\neq0\\
r=0
\end{array}\end{cases},
\end{equation}

Although the last result is unitary for any $r\neq0$, our approach regards the case $r=0$ as non-unitary, explicitly expressed as
\begin{equation}\label{nonuni}
\left(\exp\left[\kappa\pi\delta(r)\,\Delta t\right]\right)^{\dagger}\,\neq\,\left(\exp\left[\kappa\pi\delta(r)\,\Delta t\right]\right)^{-1}.
\end{equation}
despite the fact that rigorous proof has not been obtained. An important observation is that since the Taylor series expansion does not exist for generalized objects, one cannot repeat the considerations of (\ref{unibound}). Therefore, without getting into the details of how to define $\exp\left[\pi\delta(\hat{r})\,\Delta t\right]$, we can use the formal structure as a unitary operator:
\begin{equation}
\mathcal{P}(r,\,\Delta t)\,=\,\sqrt{\exp^{\dagger-1}\left[\kappa\pi\delta(r)\,\Delta t\right]\,\exp^{-1}\left[\kappa\pi\delta(r)\,\Delta t\right]}\exp\left[\kappa\pi\delta(r)\,\Delta t\right].
\end{equation}

\begin{figure}[H]
\center
\includegraphics[scale=0.5]{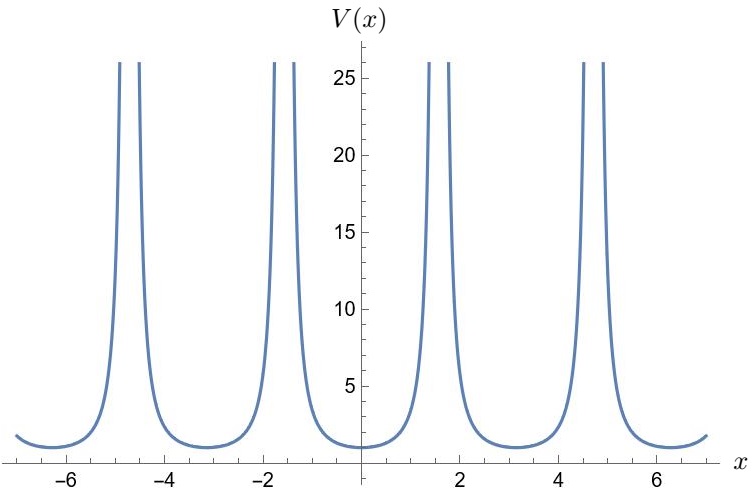}
\caption{The Poschl--Teller potential with $V_{0}=\alpha=1$.}\label{exam2del}
\end{figure}

The situation becomes more interesting when passing from potentials with only a single singular point to the case in which a measure-$0$ set of singularities is involved. An explicit example for such a potential is the Poschl--Teller potential (see Figure~\ref{exam2del}),
\begin{equation}
V(r)\,=\,\frac{V_{0}}{\cos^{2}\alpha r},
\end{equation}
which after extracting the generalized part can be written as
\begin{equation}
\lim_{\epsilon\rightarrow0}\frac{V_{0}}{\cos^{2}\alpha r-i\epsilon}\,=\,P.v.\left(\frac{V_{0}}{\cos^{2}\alpha r}\right)+i\pi\delta(\cos^{2}\alpha r).
\end{equation}

The issue with unitarity occurs at the tunneling points between regions. Without solving the singular dynamics, one has to specify it only for a specific branch of the potential and cannot realize the entire dynamics that include the singular locations. 
\vspace{-3pt}

\subsection{The Dirac Comb Potential}\label{combi}
\begin{figure}[H]
\center
\includegraphics[scale=0.11]{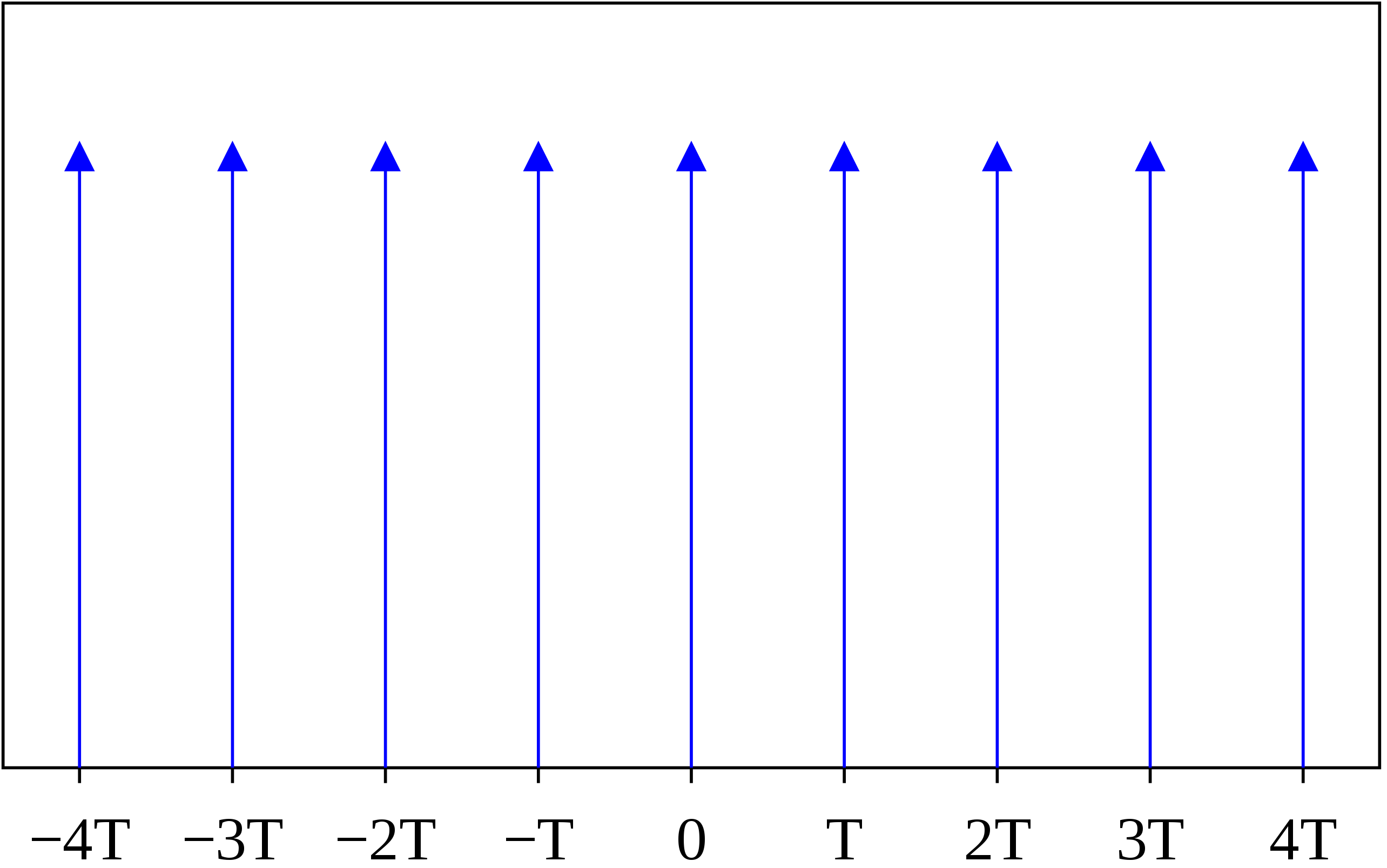}
\caption{The Dirac comb~potential.}\label{dircomhm}
\end{figure}

\begin{figure}[H]\center
\includegraphics[scale=0.62]{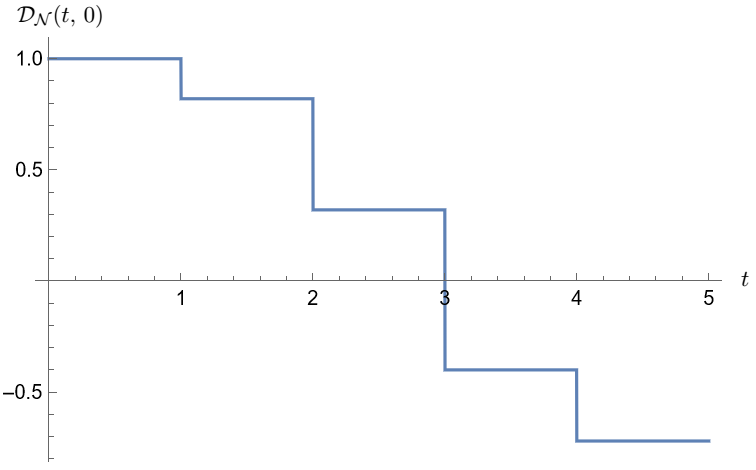}
\caption{The value of the defined part of $\mathcal{N}(t,\,0)$ for the Dirac comb Hamiltonian (\ref{dcomb}) with $n=4$ and $V_{i}\,=\,\{0.6,\,1,\,1.2,\,0.8\}$.}\label{dimb}
\end{figure}

\textls[-12]{One of the simplest examples in order to demonstrate the dynamics of a time-dependent unbounded (distributional) Hamiltonian is to use the Dirac comb potential (see \mbox{Figure~\ref{dircomhm}}). The properties of this potential has already been briefly mentioned in \mbox{Section~\ref{unbounham}}}. For our purpose, it will be sufficient to even use a truncated version of the series,
\begin{equation}\label{dcomb}
V(t)\,=\,\sum_{i=1}^{n}V_{i}\,\delta(t-t_{i}),
\end{equation}
where $V_{i}\in\mathbb{R}$ denotes an array of $n$ values. Clearly, the unitarity argument presented in  Section~\ref{notalw} is inadequate for the above choice of Hamiltonian due to the failure of additivity at the singular time $t_{i}$. Let us now examine the resulting expansion from the Dyson series. With the aid of the integration of Equation~(\ref{heavi}), one finds the following expressions for Equations~(\ref{unita}) and (\ref{exppnor}):
\begin{equation}\begin{split}\label{expnsions}
&U(t,\,0)\,=\,1\,-\,i\sum_{i=1}^{n}V_{i}\,\Theta(t-t_{i})\,-\,\sum_{i=1}^{n}V_{i}^{2}\,\int_{0}^{t}dt^{\prime}\,\delta(t^{\prime}-t_{i})\,\Theta(t^{\prime}-t_{i})\,+\,\ldots\\
&\mathcal{N}(t,\,0)\,=\,1\,-\,\frac{1}{2}\left(\sum_{i=1}^{n}V_{i}\,\Theta(t-t_{i})\right)^{2}\,+\,\sum_{i=1}^{n}V_{i}^{2}\,\int_{0}^{t}dt^{\prime}\,\delta(t^{\prime}-t_{i})\,\Theta(t^{\prime}-t_{i})\,+\,\ldots
\end{split}\end{equation}
As expected, based on the discussion in Appendix \ref{iterative}, both the results for $U$ and $\mathcal{N}$ above are ill-defined due to the last term (which cancels them when taking their products.) 
Therefore, one can think of $\mathcal{N}$ as a regulator which replaces the indefinite integrals involved in $U$ with valid integrals describing the singular dynamics. In order to gain an impression of how $\mathcal{N}$ evolves, it is useful to work with a truncated version of $\mathcal{N}$ which does not involve integration over the product of distributions,
\begin{equation}
\left.\mathcal{N}(t,\,0)\right|_{trunc.}\,\equiv\,1\,-\,\frac{1}{2}\left(\sum_{i=1}^{n}V_{i}\,\Theta(t-t_{i})\right)^{2}\,+\,\ldots
\end{equation}
\indent A plot of the last expression is shown in Figure~\ref{dimb}. We note that if $\mathcal{N}$ was differentiable like an ordinary function, then indeed $\frac{d\mathcal{N}}{dt}=0$; however, $\mathcal{N}$ is non-differentiable. The construction defined by $\mathcal{P}(t,\,0)$, as demonstrated by Equation~(\ref{pitarg2}), involves the action of $\mathcal{N}(t,\,0)$ on $U(t,\,0)$ before the evaluation of the integrals. One arrives at a result, in which the ill-defined terms canceled,
\begin{equation}
\mathcal{P}(t,\,0)\,=\,\mathcal{N}(t,\,0)\,U(t,\,0)\,=\,1\,-\,i\sum_{i=1}^{n}V_{i}\,\Theta(t-t_{i})\,-\,\frac{1}{2}\left(\sum_{i=1}^{n}V_{i}\,\Theta(t-t_{i})\right)^{2}\,+\,\ldots
\end{equation}
\indent Although our formalism does not describe what happens at the isolated transition time $t_{1}$, this is not of any concern. Practically, the time that a perturbation is turned on is well separated from the time that a measurement is conducted.
\textit{What is the problem with smearing singularities?}

Clearly, an approach that can handle distributions directly has an advantage over one that cannot. However, it is tempting to think that the problem of dealing with distributions is artificial and can be solved by replacing the original singular Hamiltonian with one that mimics it but contains no singularity. In order to carry this idea, one replaces $H(t)\rightarrow H_{\epsilon}(t)$ and smears the original singularity on a finite width $\epsilon$, which is supposed to be subsequently taken to $0$. For example, in the case of Dirac's delta function, this can be achieved by using either the nascent or Gaussian representations,
\begin{equation}\label{delt}
\delta(x)\,=\,\lim_{\epsilon\rightarrow0}\frac{\epsilon}{\pi(x^{2}+\epsilon^{2})},\qquad\qquad\delta(x)\,=\,\lim_{\epsilon\rightarrow0}\frac{1}{2\sqrt{\pi\epsilon}}\exp\left(-\frac{x^{2}}{4\epsilon}\right).
\end{equation}

Thus, a practical calculation will be performed by first replacing an originally singular integration according to the modification
\begin{equation}
\int_{0}^{t}dt^{\prime}\,H(t^{\prime})\,\rightarrow\,\int_{0}^{t}dt^{\prime}\,\lim_{\epsilon\rightarrow0}H_{\epsilon}(t^{\prime})\,\rightarrow\,\lim_{\epsilon\rightarrow0}\int_{0}^{t}dt^{\prime}\,H_{\epsilon}(t^{\prime}).
\end{equation}

The last transition involving the exchange of a limiting component 
with an integration deserves a closer look. The theorem that allows to establish the last equivalence,
\begin{equation}\label{domcon}
\int_{0}^{t}dt^{\prime}\,\lim_{\epsilon\rightarrow0}H_{\epsilon}(t^{\prime})\,=\,\lim_{\epsilon\rightarrow0}\int_{0}^{t}dt^{\prime}\,H_{\epsilon}(t^{\prime}),
\end{equation}
is known as Lebesgue’s dominated convergence~\cite{Measure1,Measure2,Measure3,Measure4}. One of its necessary conditions is to find an integrable dominating function, satisfying $\left|H_{\epsilon}(t^{\prime})\right|<\left|H(t^{\prime})\right|$ for any value of $\epsilon$ and $t^{\prime}\in[0,t]$. Although it is true that, sometimes, one can perform transition (\ref{domcon}) even without satisfying the necessary condition, that is not the case in general. Several examples can be found in the literature of measure theory. For~example, let us define the following:
\begin{equation}
f_{n}(x)\,=\,\begin{cases}
\begin{array}{c}
\frac{1}{n}\\
0
\end{array} & \begin{array}{c}
0<x<n\\
x\leq0;\;x\geq n
\end{array}\end{cases}.
\end{equation}

Then, since $\lim_{n\rightarrow0}f_{n}(x)\,=\,0$, then we obtain
\begin{equation}
0\,=\,\int_{-\infty}^{\infty}dx\,\lim_{n\rightarrow0}f_{n}(x)\,\neq\,\lim_{n\rightarrow0}\int_{-\infty}^{\infty}dx\,f_{n}(x)\,=\,1.
\end{equation}

\textls[-25]{An additional familiar example is given by $f_{n}(x)=nxe^{-nx^{2}}$, for~which \mbox{$\lim_{n\rightarrow\infty}f_{n}(x)=0$,}} then
\begin{equation}
0\,=\,\int_{-\infty}^{\infty}dx\,\lim_{n\rightarrow\infty}f_{n}(x)\,\neq\,\lim_{n\rightarrow\infty}\int_{-\infty}^{\infty}dx\,f_{n}(x)\,=\,\frac{1}{2}.
\end{equation}

In fact, according to Fatou's lemma, the inequality between the results holds in general:
\begin{equation}
\int dx\,\lim_{n\rightarrow0}f_{n}(x)\,\leq\,\lim_{n\rightarrow0}\inf\int dx\,f_{n}(x).
\end{equation}

Therefore, the only mathematically consistent way to perform the smearing trick is when the width is kept as a finite parameter. That means that a smearing parameter ends up taking a physical role in the formulation which demands that they be provided for a meaningful interpretation. 
In gauge theories, the Hamiltonian is often expressed via fields, which not only makes such an implementation inconvenient but also breaks the gauge invariance. It can also be shown that the ``smearing trick'' becomes less applicable once higher-order terms are considered. Based on (\ref{delt}), assuming $H(t)=\delta(t-t_{1})$ and wrongly exchanged the limit and integration, the second-order term of (\ref{unita}) becomes
\begin{equation}
\int_{0}^{t}dt^{\prime}\,H(t^{\prime})\int_{0}^{t^{\prime}}dt^{\prime\prime}\,H(t^{\prime\prime})\,\rightarrow\,\lim_{\epsilon_{1,2\rightarrow0}}\int_{0}^{t}dt^{\prime}\,\frac{1}{2\sqrt{\pi\epsilon_{2}}}e^{-\frac{(t^{\prime}-t_{1})^{2}}{4\epsilon_{2}}}\int_{0}^{t^{\prime}}dt^{\prime\prime}\,\frac{1}{2\sqrt{\pi\epsilon_{1}}}e^{-\frac{(t^{\prime\prime}-t_{1})^{2}}{4\epsilon_{1}}}.
\end{equation}
\indent Such a result is not only difficult to compute but also crucially dependent on the way in which the limit $\epsilon_{1,2}\rightarrow0$ is taken.

\section{Conclusions}\label{conclu}
In this part, a summarized list of the consequences is provided for adopting (\ref{pitaron}) as the solution for the Schrödinger equation. These implications affect both the perturbative and the non-perturbative aspects.
\begin{itemize}
\item[$\circledast$] Unitarity is the key property of the quantum dynamics. The currently widely used solution $\hat{U}$ should be regarded as a unitary operator only when working with a self-adjoint Hamiltonian on a finite dimensional Hilbert space (each state has a norm given as a finite number). For infinitely dimensional Hilbert spaces, Hermiticity and self-adjoint are not equivalent properties, and Hermiticity is an insufficient condition to generate a unitary evolution.
\item[$\circledast$] For the proposed solution, unitarity
is manifestly maintained at all orders and at any given moment of the evolution rather than asymptotically. The ordinary probabilistic interpretation of Born is maintained for any Hamiltonian, even in the cases of non-Hermitian and unbounded Hamiltonians.  The Liouville part of the evolution generated by $\hat{\mathcal{N}}$ acts as a ``dynamical probability conservation operator''. The action of this operator produces the correction for the occurrences of unitarity breaking which is involved in the dynamics of $\hat{U}$.
\item[$\circledast$] The solution provided will hopefully pave the way for a better understanding of the various quantum systems in which unitarity is currently assumed to be absent, including systems with a non-Hermitian Hamiltonians, field theories on non-commutative spaces~\cite{Gomis1,Gomis2,Gomis3}, field theories on factional dimensions~\cite{Rychkov1,Rychkov2}, or~open quantum systems~\cite{open1,open2,open3,open4}.
The new solution becomes crucially important when studying the time evolution of higher dimensional entangled states (such as biphotons). These states live on a product of two infinitely dimensional spaces, $\mathcal{H}_{a}\otimes\mathcal{H}_{b}$ with $\dim\left[\mathcal{H}_{a/b}\right]=\infty$,
\begin{equation}
\left|\psi\right\rangle \,=\,\int dx_{a,b}\,f(x_{a},\,x_{b})\left|x_{a}\right\rangle \otimes\left|x_{b}\right\rangle. 
\end{equation}
One can show that even if such a state is initially prepared as normalized, that will be spoiled as time goes by if the Hamiltonian is unbounded, unless we introduce $\hat{\mathcal{N}}$.
The analysis of these states is a potential option for the experimental verification of our proposed idea.
\item[$\circledast$] The reason which initiated this study and was not mentioned is the investigation of the scattering amplitudes for entangled states. In this setup, unlike where fully on-shell states are involved, the JIMWLK equation~\cite{jimwlk} is, strangely, no longer applicable in the way expected based on~\cite{iancu1,iancu2}. That happens due to the normalization of the WF via a simple overall $\mathcal{Z}$ factor~\cite{adhoc1,adhoc2} breakdown and a convolution is required instead. It is interesting to point out that based on logic alone, the analysis in \cite{Dumitru} has been conducted in a way that is consistent with this paper and cannot be performed~otherwise.
\end{itemize}

\vspace{3pt}
\funding{This research was supported by Chinese academy of sciences president's international fellowship initiative, grant no. 2025PVC0019.}

\dataavailability{The original contributions presented in the study are included in the article, further inquiries can be directed to the corresponding author.} 

\acknowledgments{First and foremost, Y.M. would like to thank his mother for which this paper is dedicated. Y.M. would like to greatly thank A. Ramallo for their many valuable discussions, as well as their guidance and support. Y.M also thanks E. Iancu, A. Dumitru, E. Gandelman, D. Cohen, G. Beuf, F. Salazar, F. Perez, A. H. Mueller, E. Buks, O. M. Shalit, J. Lampart, A. Jadczyk, and various other people with whom they had the chance to discuss about this project. Y.M. is also grateful for the Mathematics Stack Exchange community for providing help with regard to the technical aspects involved in this project. A very special thanks goes to T. Lappi from the University of Jyväskylä, where this project was first initiated, for plenty of discussions on this matter in a way that eventually shaped the structure of this paper. As a matter of fact, this paper evolved through the process of striving to obtain answers for his various interesting questions that arose along the way.}

\conflictsofinterest{The author has no relevant financial or non-financial conflicts of interest to~disclose.}

\appendixtitles{yes}
\appendixstart
\appendix

\section{Why Might the Iterative Method Fail?}\label{iterative}
In this section, we would like to discuss Picard's method of successive approximations~\cite{Lindelof,Evolutiona} in order to analyze the capacity of this approach to yield the solution for a differential evolution equation of first order. Our intention is to consider a differential equation of the type:
\begin{equation}\label{orddiff}
\frac{dy}{dx}\,=\,f(x,y)\:,
\end{equation}
subject to the initial condition $y(x_{0})=y_{0}$. The key point of Picard's idea consists in replacing the original differential equation with a system of difference equations that approximate the evolution occurring in a small neighborhood of the domain, such that (\ref{orddiff}) is essentially replaced by solving the system $y_{1}-y_{0}\,\approx\,f(x,y_{0})\,\Delta x,\:y_{2}-y_{1}\,\approx\,f(x,y_{1})\,\Delta x,\:\ldots,\:y_{n+1}-y_{n}\,\approx\,f(x,y_{n})\,\Delta x$, leading to $y_{n+1}-y_{0}\,\approx\,f(x,y_{0})\,\Delta x+f(x,y_{0}+f(x,y_{0})\,\Delta x)\,\Delta x+\ldots$ \cite{lax}. As demonstrated below, under certain assumptions, the resulting difference equation can reliably approximate the evolution on the whole domain. In that case, the resulting solution uniquely converges to the analytic solution of the original equation.  Our intention is to provide an answer for the fundamental questions: \textit{What are the necessary conditions for the iterative method to be applicable? Is the discretization limit always admissible?}

Let us start by applying the method in the non-negative region $[0,\,x]$, with the simple choice of $f(x,y)\,=\,gy$ with $y_{0}=1$. The analytical solution corresponding to our choice can naturally be found by direct integration, $y(x)=e^{gx}$. Now, let us reproduce the last solution via the iterative method. Our initial assumption is that the differential equation above allows us to start a process of successive approximations by rewriting it in the integral form~as
\begin{equation}
y_{n+1}(x)\,=\,y_{0}+\int_{0}^{x}dx^{\prime}\,f(x^{\prime},y_{n}(x^{\prime})).
\end{equation}
\indent After a few iterations, the following results are obtained:
\begin{equation}\begin{split}
&y_{1}(x)\,=\,1+g\int_{0}^{x}1\,dx^{\prime}=1+gx,\\
&y_{2}(x)\,=\,1+g\int_{0}^{x}(1+gx)\,dx^{\prime}=1+gx+\frac{1}{2!}g^{2}x^{2},\\
&y_{3}(x)\,=\,1+g\int_{0}^{x}(1+gx+\frac{1}{2!}g^{2}x^{2})\,dx^{\prime}=1+x+\frac{1}{2!}g^{2}x^{2}+\frac{1}{3!}g^{3}x^{3},
\end{split}\end{equation}
from which we observe the series that is generated after $n$ iterations,
\begin{equation}
y_{n}(x)\,=\,\sum_{i=0}^{n}\frac{g^{n}}{n!}x^{n}.
\end{equation}

As expected, the sequence $\{y_{n}(x)\}$ converges to the Taylor series for the exponent,
\begin{equation}
\lim_{n\rightarrow\infty}y_{n}(x)\,=\,e^{gx}\,,
\end{equation}
which indeed reproduces the expected result. This last equivalence was, in fact, guaranteed to us by the Picard–Lindelöf theorem~\cite{Evolutiona}. As discussed at length in~\cite{byron}, the error at the $n$th step of the iterative procedure in the interval $x_{0}-h<x<x_{0}+h$ with $h\,\equiv\,\min\left(a,\,\frac{b}{M}\right)$, where $\left|x-x_{0}\right|\,\leq\,a$ and $\left|y(x)-y_{0}\right|\,\leq\,b$, is bounded by the inequality
\begin{equation}
\left|y_{n}(x)-y(x)\right|\,\leq\,\frac{MN^{n-1}}{n!}h^{n},
\end{equation}
with the definitions $M\,\equiv\,\max_{(x,y)\in\mathcal{D}}\left|f(x,y)\right|$, $N\,\equiv\,\max_{(x,y)\in\mathcal{D}}\left|\frac{\partial f}{\partial y}\right|$\,. Clearly, in order for the solution obtained by the iterative method to be considered valid, it must converge asymptotically with the analytic solution,
\begin{equation}
\lim_{n\rightarrow\infty}\left|y_{n}(x)-y(x)\right|\,=\,0.
\end{equation}
\indent However, as we are about to see, the above theorem will fail to imply uniqueness when singularities are involved. For example, let us take a closer look at what happens when trying to solve a differential equation with RHS involving Dirac's delta function~\cite{distrib1,distrib2,distrib3},
\begin{equation}
\frac{df(x)}{dx}\,=\,\delta(x-a)\,f(x),
\end{equation}
subject to the initial condition $f(0)=1$ with $a>0$. The direct solution method leads to $f(x)\,=\,e^{\Theta(x-a)}$, where we used the defining relation for the Heaveside theta function. Generally, if~the signs of $a$ and $x$ are unknown, $\int_{0}^{x}f(x^{\prime})\,\delta(x^{\prime}-a)\,dx^{\prime}\,=\,f(a)\,(2\Theta(x)-1)\,\Theta(a-x\Theta(-x))\,\Theta(-a+x\Theta(x))$.
\begin{equation}\label{heavi}
\int_{0}^{x}dx^{\prime}\,\delta(x^{\prime}-a)\,=\,\Theta(x-a).
\end{equation}

Now, let us try to reproduce the last result by the iterative method. For that purpose, we present the solution as
\begin{equation}
f(x)\,=\,1+\int_{0}^{x}dx^{\prime}\,\delta(x^{\prime}-a)\,f(x^{\prime})\,,
\end{equation}
and then by substitution, we obtain
\begin{equation}\label{iterdelta}
f(x)\,=\,1+\int_{0}^{x}dx^{\prime}\,\delta(x^{\prime}-a)\left[1+\int_{0}^{x^{\prime}}dx^{\prime\prime}\,\delta(x^{\prime\prime}-a)\,f(x^{\prime\prime})\right]\,.
\end{equation}

After performing the integration of the inner brackets by using (\ref{heavi}), it is clear that something goes wrong. We arrive at a badly defined result that involves integration over a product of two distributions~\cite{Schwarz},
\begin{equation}\label{undefi}
\int_{0}^{x}dx^{\prime}\,\delta(x^{\prime}-a)\,\Theta(x^{\prime}-a)\;\longrightarrow\;\mathtt{Indefinite}.
\end{equation}
\indent It is therefore clear that the iterative series (\ref{iterdelta}) will not converge at the expected result. We realize that in order for the iterative method to work, the integrand has to be well defined (measurable function) at each part of the domain. A more practical example, which allows us to get a deeper understanding of what happens when the conditions for Picard's theorem fails, is given by the differential equation
\begin{equation}\label{singdiff}
x\frac{dy}{dx}\,=\,A,
\end{equation}
which has an unbounded RHS when brought to the form of (\ref{orddiff}). The solution is often taken as $y(x)\,=\,A\log|x|+C$, although~a more general discontinuous solution can be~found:
\begin{equation}
y\,=\,\begin{cases}
\begin{array}{c}
A\log(x)+C\\
A\log(-x)+B
\end{array} & \begin{array}{c}
x>0\\
x<0.
\end{array}\end{cases}
\end{equation}
\indent \textls[-15]{In terms of distributions that can be expressed as $y(x)\,=\,A\,\log|x|+(B-C)\,\Theta(-x)+C$.} Note that even though the constant term changes in the passage between the two disjoint regions, this is still a valid solution. The discontinuous part or, alternatively, the singular dynamic at $x=0$, is exactly where the iterative procedure fails to provide a definitive answer. Thus, our main conclusion in this part is that singular differential equations, which are typically of the form $x^{n}\frac{d^{m}y}{dx^{m}}\,=\,0$ with $n,m\in\mathbb{N}$ cannot be solved by a solution method that is based solely on the iterative method. For example, as discussed in~\cite{Kan1,Kan2}, if~$m=1$, the obtained result is given by $y(x)\,=\,c_{1}\,+\,c_{2}\Theta(x)\,+\,c_{3}\delta(x)\,+\,c_{4}\delta^{\prime}(x)\,+\,\ldots\,+\,c_{n+1}\delta^{(n-2)}(x)$, for which the approximate expressions cannot be found.

\subsection*{The Uniqueness of $\hat{\mathcal{N}}$}
The operator inside the square root of Equation~(\ref{definorm}) is a positive definite operator. As such, its square has a unique root given by a positive definite operator~\cite{matana1} that seems natural to regard as ``physical''. However, it is true that additional, non-positive roots can be found as well. Such a choice was, in fact, already involved in (\ref{nonher}), as demonstrated by:
\begin{equation}
\hat{\mathcal{N}}(t,t_{0})\,=\,\sqrt{\hat{\boldsymbol{1}}}\exp\left(\hat{\mathcal{J}}\Delta t\right),
\end{equation}
where we took the positive root option, $\sqrt{\hat{\boldsymbol{1}}}\,=\,\left(\begin{array}{cc}
1 & 0\\
0 & 1
\end{array}\right)$.
In general, the~identity matrix of dimension $2$ has infinitely many square roots given by
\begin{equation}
\sqrt{\left(\begin{array}{cc}
1 & 0\\
0 & 1
\end{array}\right)}\,=\,\left(\begin{array}{cc}
\pm1 & 0\\
0 & \pm1
\end{array}\right)\quad and\quad\left(\begin{array}{cc}
a & b\\
c & -a
\end{array}\right).
\end{equation}
with $a^{2}+bc=1$.  The~collection of all these roots are related with each other via a fixed unitary transformation, $\sqrt{\hat{\boldsymbol{1}}}\,=\,\hat{U}\,\hat{\boldsymbol{1}}$. While the importance of these additional solutions deserves an interpretation, in various situations, this will eventually have no effect on the actual observables (as an overall phase). It should also be mentioned that from the mathematical side, the non-uniqueness of the solution is, in fact, not surprising. The theorem that ensures the uniqueness of the differential Equation (\ref{orddiff}) is the Picard–Lindelöf (PL) theorem. This theorem is based on the continuity of $f(x,y)\in\mathbb{R}$ for generating the final result via successive iteration procedures. In order for the method to work, it is required that a finite bound operator exists inside the entire considered domain,
\begin{equation}\label{req}
\left|f(x,y)\right|,\,\left|\frac{\partial f(x,y)}{\partial y}\right|\,<\,\infty.
\end{equation}
\indent In the cases where $f(x,y)\in\mathbb{C}$, or if the above condition does not hold, there are counterexamples showing that uniqueness does not hold as well. Therefore, the solution should be expected to be unique only for the regime in which the underlying assumptions for the PL theorem are satisfied and, in that case, indeed $\hat{\mathcal{N}}=\hat{\boldsymbol{1}}$. Outside this regime, there is a priori no reason to expect uniqueness to continue to hold, as a feature that typically characterizes non-Hermitian systems.

\section{The Dynamics of \boldmath{$\hat{\mathcal{N}}$}}\label{dynamn}
In this part, we provide the proof for the second equation of (\ref{twoeqs}). From the defining relation for the normalization operator, $\hat{U}^{\dagger}(t,\,t_{0})\,\mathcal{\hat{N}}^{2}(t,\,t_{0})\,\hat{U}(t,\,t_{0})\,=\,\mathbf{\hat{1}}$. By~applying the time derivative on the two sides of this equation we obtain:
\begin{equation}\begin{split}
&\hat{U}^{\dagger}(t,\,t_{0})\,\left(\mathcal{\hat{N}}(t,\,t_{0})\frac{d\mathcal{\hat{N}}(t,\,t_{0})}{dt}\,+\,\frac{d\mathcal{\hat{N}}(t,\,t_{0})}{dt}\mathcal{\hat{N}}(t,\,t_{0})\right)\,\hat{U}(t,\,t_{0})\\
&=\,-\frac{d\hat{U}^{\dagger}(t,\,t_{0})}{dt}\,\mathcal{\hat{N}}^{2}(t,\,t_{0})\,\hat{U}(t,\,t_{0})\,-\,\hat{U}^{\dagger}(t,\,t_{0})\,\mathcal{\hat{N}}^{2}(t,\,t_{0})\,\frac{d\hat{U}(t,\,t_{0})}{dt}.
\end{split}\end{equation}

After multiplication by $\hat{U}^{\dagger-1}(t,\,t_{0})$ from the right side and $\hat{U}^{-1}(t,\,t_{0})$ from the left~side:
\begin{equation}\begin{split}
&\mathcal{\hat{N}}(t,\,t_{0})\frac{d\mathcal{\hat{N}}(t,\,t_{0})}{dt}\,+\,\frac{d\mathcal{\hat{N}}(t,\,t_{0})}{dt}\mathcal{\hat{N}}(t,\,t_{0})\\
&=\,-\hat{U}^{\dagger-1}(t,\,t_{0})\,\frac{d\hat{U}^{\dagger}(t,\,t_{0})}{dt}\,\mathcal{\hat{N}}^{2}(t,\,t_{0})\,-\,\mathcal{\hat{N}}^{2}(t,\,t_{0})\,\frac{d\hat{U}(t,\,t_{0})}{dt}\,\hat{U}^{-1}(t,\,t_{0}),
\end{split}\end{equation}
which is a Sylvester-type equation \cite{Sylvester1,Sylvester2} for $\frac{d\mathcal{\hat{N}}(t,\,t_{0})}{dt}$. In our case, due to the positive definite property of $\mathcal{\hat{N}}$, this equation can be solved in a unique manner. In fact, this equation is not just a Sylvester equation \cite{Sylvester1,Sylvester2}, but more precisely, the continuous time Lyapunov equation. These are equations of the type $\mathbf{\hat{A}}\mathbf{\hat{X}}\,+\,\mathbf{\hat{X}}\mathbf{\hat{A}}^{\dagger}\,+\,\mathbf{\hat{Q}}\,=\,0$, where $\mathbf{\hat{Q}}$ is a self-adjoint operator \cite{Lyapunov}. As~such, its solution is known to be given explicitly by
\begin{equation}\begin{split}
&\frac{d\mathcal{\hat{N}}(t,\,t_{0})}{dt}\,=\,-\int_{0}^{\infty}dy\,e^{-y\mathcal{\hat{N}}}\\
&\times\left(\hat{U}^{\dagger-1}(t,\,t_{0})\,\frac{d\hat{U}^{\dagger}(t,\,t_{0})}{dt}\,\mathcal{\hat{N}}^{2}(t,\,t_{0})\,+\,\mathcal{\hat{N}}^{2}(t,\,t_{0})\,\frac{d\hat{U}(t,\,t_{0})}{dt}\,\hat{U}^{-1}(t,\,t_{0})\right)e^{-y\mathcal{\hat{N}}}.
\end{split}\end{equation}

Introducing the dynamics of $\hat{U}$ from Equation~(\ref{twoeqs}) into the above result we obtain:
\begin{equation}\begin{split}\label{smla}
&\frac{d\mathcal{\hat{N}}(t,\,t_{0})}{dt}\,=\,-\int_{0}^{\infty}dy\,e^{-y\mathcal{\hat{N}}}\\
&\times\left(i\left(\hat{\mathcal{H}}(t)+i\mathcal{\hat{J}}(t)\right)\,\mathcal{\hat{N}}^{2}(t,\,t_{0})\,-\,i\mathcal{\hat{N}}^{2}(t,\,t_{0})\,\left(\hat{\mathcal{H}}(t)-i\mathcal{\hat{J}}(t)\right)\right)e^{-y\mathcal{\hat{N}}}.
\end{split}\end{equation}

At this point one can simplify further by using the identity 
\begin{equation}
\left[\hat{\mathcal{H}}(t),\,\mathcal{\hat{N}}^{2}(t,\,t_{0})\right]=\mathcal{\hat{N}}(t,\,t_{0})\left[\hat{\mathcal{H}}(t),\,\mathcal{\hat{N}}(t,\,t_{0})\right]\,+\,\left[\hat{\mathcal{H}}(t),\,\mathcal{\hat{N}}(t,\,t_{0})\right]\mathcal{\hat{N}}(t,\,t_{0}),
\end{equation}
along with the observation that since the operator $\hat{\mathcal{N}}$ is fully expressible based on the operator $\hat{\mathcal{J}}$ since they share a common basis and domain, 
\begin{equation}
\left[\mathcal{\hat{N}}(t,\,t_{0}),\,\mathcal{\hat{J}}(t)\right]=0.
\end{equation}

At last, Equation (\ref{smla}) is rewritten as
\vspace{-6pt}
\begin{equation}
\frac{d\mathcal{\hat{N}}(t,\,t_{0})}{dt}=-\int_{0}^{\infty}dy\,\frac{d}{dy}\left[e^{-y\mathcal{\hat{N}}}\left(-i\left[\hat{\mathcal{H}}(t),\mathcal{\hat{N}}(t,t_{0})\right]+\mathcal{\hat{N}}(t,\,t_{0})\,\mathcal{\hat{J}}(t)\right)e^{-y\mathcal{\hat{N}}}\right].
\end{equation}

Finally, based on an immediate integration,
\begin{equation}
\int_{0}^{\infty}dy\frac{d}{dy}\left[e^{-y\hat{\mathcal{N}}}\hat{\mathcal{O}}(t,\,t_{0})e^{-y\hat{\mathcal{N}}}\right]\,=\,\left.e^{-y\hat{\mathcal{N}}}\hat{\mathcal{O}}(t,\,t_{0})e^{-y\hat{\mathcal{N}}}\right|_{0}^{\infty}=-\hat{\mathcal{O}}(t,\,t_{0}),
\end{equation}
one establishes the fact that $\mathcal{\hat{N}}(t,\,t_{0})$ has a non-trivial dynamics that is given by the second equation written in (\ref{twoeqs}).

\reftitle{References}

\begin{adjustwidth}{-\extralength}{0cm}

\PublishersNote{}
\end{adjustwidth}

\begin{thebibliography}{999}
\bibitem{Schro}
Schrödinger, E. An undulatory theory of the mechanics of atoms and molecules. \emph{Phys. Rev.} \textbf{1926}, \emph{28}, 1049. [\href{http://doi.org/10.1103/PhysRev.28.1049}{CrossRef}]

\bibitem{landau1}
Landau, L.D.;  Lifshitz, E.M. \emph{Quantum Mechanics: Non-Relativistic Theory};  Elsevier:  Amsterdam, The Netherlands, 2013; Volume 3.

\bibitem{landau2}
Böhm, A.  \emph{Quantum Mechanics: Foundations and Applications}; Springer Science \& Business Media: Berlin/Heidelberg, Germany,~2013.

\bibitem{landau3}
Dirac, P.A.M.  \emph{The Principles of Quantum Mechanics (No. 27)}; Oxford University Press: Oxford, UK, 1981.

\bibitem{landau4}
Shankar, R.  \emph{Principles of Quantum Mechanics}; Springer Science \& Business Media:  Berlin/Heidelberg, Germany, 2012.

\bibitem{landau5}
Sakurai, J.J.; Commins, E.D. \emph{Modern Quantum Mechanics}, revised ed.; Addison Wesley: Boston, MA, USA, 1995.

\bibitem{Peskin1}
Peskin, M.E. \emph{An Introduction to Quantum Field Theory}; CRC Press:  Boca Raton, FL, USA, 2018.

\bibitem{Peskin2}
Kaiser, D. \emph{Lectures of Sidney Coleman on Quantum Field Theory: Foreword by David Kaiser}; World Scientific Publishing: Singapore,~2018.

\bibitem{Peskin3}
Itzykson, C.; Zuber, J.B.  \emph{Quantum Field Theory}; Courier Corporation: Chelmsford, MA, USA, 2012.

\bibitem{Peskin4}
Schwartz, M.D.  \emph{Quantum Field Theory and the Standard Model}; Cambridge University Press: Cambridge, UK, 2014.

\bibitem{func1}
Yosida, K.  \emph{Functional Analysis}; Springer Science \& Business Media:  Berlin/Heidelberg, Germany, 2012.

\bibitem{func2}
Suhubi, E.  \emph{Functional Analysis}; Springer Science \& Business Media: Berlin/Heidelberg, Germany, 2013.

\bibitem{fu1}
Reed, M. \emph{Methods of Modern Mathematical Physics: Functional Analysis}; Elsevier: Amsterdam, The Netherlands, 2012.

\bibitem{fu2}
Shalit, O.M. \emph{A First Course in Functional Analysis}; CRC Press: Boca Raton, FL, USA, 2017.

\bibitem{freema1}
Dyson, F.J. The radiation theories of Tomonaga, Schwinger, and Feynman. \emph{Phys. Rev.} \textbf{1949}, \emph{75}, 486. [\href{http://dx.doi.org/10.1103/PhysRev.75.486}{CrossRef}]

\bibitem{freema2}
Dyson, F.J. The $S$ matrix in quantum electrodynamics. \emph{Phys. Rev.} \textbf{1949}, \emph{75}, 1736. [\href{http://dx.doi.org/10.1103/PhysRev.75.1736}{CrossRef}]

\bibitem{matana1}
Horn, R.A.; Johnson, C.R.  Positive definite and semidefinite matrices. In \emph{Matrix Analysis}; Cambridge University Press: New York, NY, USA, 2013; pp. 425--515.
\bibitem{matana2}
Zhang, X.-D. \emph{Matrix Analysis and Applications}; Cambridge University Press: New York, NY, USA, 2017.
\bibitem{matana3}
Higham, N.J.  \emph{Functions of Matrices: Theory and Computation}; Society for Industrial and Applied Mathematics: Philadelphia, PA, USA, 2008.

\bibitem{Born}
Born, M. Quantenmechanik der Stosvorgange. \emph{Z. F. Phys.} \textbf{1926}, \emph{38}, 803--827. [\href{http://dx.doi.org/10.1007/BF01397184}{CrossRef}]

\bibitem{Rudin1}
Apelian, C.; Surace, S.  \emph{Real and Complex Analysis}; CRC Press:  Boca Raton, FL, USA, 2009.
\bibitem{Rudin2}
Simon, B.  \emph{Real Analysis}; American Mathematical Association: Washington, DC, USA, 2015.
\bibitem{Rudin3}
Folland, G.B.  \emph{Real Analysis: Modern Techniques and Their Applications}; John Wiley \& Sons:  Hoboken, NJ, USA, 1999; Volume 40.
\bibitem{Rudin4}
Makarov, B.; Podkorytov, A.  \emph{Real Analysis: Measures, Integrals and Applications}; Springer Science \& Business Media:  Berlin/Heidelberg, Germany, 2013.

\bibitem{Tes1}
Teschl, G. \emph{Mathematical Methods in Quantum Mechanics}; American Mathematical Association: Washington, DC, USA, 2014; \mbox{Volume 157}.
\bibitem{Tes2}
Dimock, J. \emph{Quantum Mechanics and Quantum Field Theory: A Mathematical Primer}; Cambridge University Press: Cambridge, UK,~2011.

\bibitem{unbounded1}
Schmüdgen, K. \emph{Unbounded Self-Adjoint Operators on Hilbert Space}; Springer Science Business Media: Berlin/Heidelberg, Germany, 2012; Volume 265.
\bibitem{unbounded2}
Goldberg, S.  \emph{Unbounded Linear Operators: Theory and Applications}; Courier Corporation: Chelmsford, MA, USA, 2006.

\bibitem{stone1}
Hall, B.C.  \emph{Quantum Theory for Mathematicians}; Springer Publication: Berlin/Heidelberg, Germany, 2013.
\bibitem{stone2}
Neumann, J.V.  Uber einen satz von herrn mh stone. \emph{Ann. Math.} \textbf{1932}, \emph{33}, 567--573. [\href{http://dx.doi.org/10.2307/1968535}{CrossRef}]

\bibitem{Gieres1}
Gieres, F.  Mathematical surprises and Dirac's formalism in quantum mechanics. \emph{Rep. Prog. Phys.} \textbf{2000}, \emph{63}, 1893. [\href{http://dx.doi.org/10.1088/0034-4885/63/12/201}{CrossRef}]
\bibitem{Gieres2}
Bonneau, G.; Faraut, J.; Valent, G.  Self-adjoint extensions of operators and the teaching of quantum mechanics. \emph{Am. J. Phys.} \textbf{2001}, \emph{69}, 322--331. [\href{http://dx.doi.org/10.1119/1.1328351}{CrossRef}]
\bibitem{Gieres3}
Gitman, D.M.; Tyutin, I.V.; Voronov, B.L.  \emph{Self-Adjoint Extensions in Quantum Mechanics: General Theory and Applications to Schrödinger and Dirac Equations with Singular Potentials};  Springer Science \& Business Media:  Berlin/Heidelberg, Germany, 2012; Volume 62.

\bibitem{polar1}
Hall, B.C.; Hall, B.C.  \emph{Lie Groups, Lie Algebras, and Representations}; Springer: New York, NY, USA, 2013; pp. 333--366.
\bibitem{polar2}
Lipschutz, S.; Lipson, M.L.  \emph{Schaum's Outline of Linear Algebra}; McGraw-Hill Education: New York, NY, USA, 2018.

\bibitem{Markov1}
Stroock, D.W.  \emph{An Introduction to Markov Processes}; Springer Science \& Business Media: Berlin/Heidelberg, Germany, 2013; \mbox{Volume 230}.
\bibitem{Markov2}
Dynkin, E.B.  \emph{Theory of Markov Processes}; Courier Corporation: Chelmsford, MA, USA, 2012.
\bibitem{Markov3}
Rivas, A.; Huelga, S.F.; Rivas, A.; Huelga, S.F.  Quantum Markov process: Mathematical structure. In \emph{Open Quantum Systems: An Introduction}; Springer:  Berlin/Heidelberg, Germany, 2012; pp. 33--48.

\bibitem{Trotter}
Trotter, H.F.  \emph{Approximation of Semi-Groups of Operators}; Mathematical Sciences Publishers: Berkeley, CA, USA, {1958}.

\bibitem{Flugge1}
Flugge, S.  \emph{Practical Quantum Mechanics}; Springer Science \& Business Media: Berlin/Heidelberg, Germany, 1999; Volume 177.
\bibitem{Flugge2}
Galitskii, V.M.; Karnakov, B.; Galitski, V.; Kogan, V.I.  \emph{Exploring Quantum Mechanics: A Collection of 700+ Solved Problems for Students, Lecturers, and Researchers}; Oxford University Press: Oxford, MS, USA, 2013.

\bibitem{Bogolubov}
Bogolubov, N.N., Jr. \emph{Introduction to Quantum Statistical Mechanics}; World Scientific Publishing Company: Singapore, 2009.

\bibitem{Neumann}
von Neumann, J. Wahrscheinlichkeitstheoretischer Aufbau der Quantenmechanik. \emph{Nachr. Ges. Wiss. Göttingen} \textbf{1927}, \emph{1}, 245--272.

\bibitem{Moiseyev1}
Bountis, T.; Skokos, H.  \emph{Complex Hamiltonian Dynamics}; Springer Science \& Business Media: Berlin/Heidelberg, Germany, 2012; Volume 10.
\bibitem{Moiseyev2}
Berry, M.V. Optical polarization evolution near a non-Hermitian degeneracy. \emph{J. Opt.} \textbf{2011}, \emph{13}, 11570. [\href{http://dx.doi.org/10.1088/2040-8978/13/11/115701}{CrossRef}]
\bibitem{Moiseyev3}
Berry, M.V. Physics of nonhermitian degeneracies. \emph{Czech. J. Phys.} \textbf{2004}, 54, 1039--1047. [\href{http://dx.doi.org/10.1023/B:CJOP.0000044002.05657.04}{CrossRef}]
\bibitem{Moiseyev4}
Bender, C.M. Making sense of non-Hermitian Hamiltonians. \emph{Rep. Prog. Phys.} \textbf{2007}, \emph{70}, 947. [\href{http://dx.doi.org/10.1088/0034-4885/70/6/R03}{CrossRef}]
\bibitem{Moiseyev5}
Moiseyev, N. \emph{Non-Hermitian Quantum Mechanics}; Cambridge University Press: Cambridge, UK, 2011.

\bibitem{Moiseyev6}
\textls[-35]{Mostafazadeh, A. Pseudo-Hermitian representation of quantum mechanics. \emph{Int. J. Geom. Methods Mod. Phys.} \textbf{2010}, \emph{7}, 1191--1306. [\href{http://dx.doi.org/10.1142/S0219887810004816}{CrossRef}]}

\bibitem{Moiseyev7}
Ashida, Y.;  Gong, Z.; Ueda, M.  Non-hermitian physics. \emph{Adv. Phys.} \textbf{2020}, \emph{69}, 249--435.

\bibitem{Hilbert1}
Prugovecki, E.  \emph{Quantum Mechanics in Hilbert Space}; Academic Press: Cambridge, MA, USA, 1982.
\bibitem{Hilbert2}
Gallone, F.  \emph{Hilbert Space and Quantum Mechanics}; World Scientific Publishing Company: Singapore, 2014.

\bibitem{Berezanski}
Berezanskiĭ, I.M.  \emph{Expansions in Eigenfunctions of Selfadjoint Operators}; American Mathematical Association: Washington, DC, USA, 1968; Volume 17. 

 \bibitem{Uwe1}
Mariani, A.; Wiese, U.J. Self-adjoint Momentum Operator for a Particle Confined in a Multi Dimensional Cavity. \emph{J. Math. Phys.} \textbf{2023}, \emph{65}, 042102. [\href{http://dx.doi.org/10.1063/5.0178419}{CrossRef}]

\bibitem{Uwe2}
Al-Hashimi, M.H.; Wiese, U.-J. Alternative momentum concept for a quantum mechanical particle in a box. \emph{Phys. Rev. Res.} \textbf{2021},  \emph{3},  L042008. [\href{http://dx.doi.org/10.1063/5.0178419}{CrossRef}]

\bibitem{Moretti}
Moretti, V.  \emph{Spectral Theory and Quantum Mechanics}; UNITEXT, Italy; Springer: Cham, Switzerland, 2017. [\href{http://dx.doi.org/10.1103/PhysRevResearch.3.L042008}{CrossRef}]

\bibitem{Solvable}
Albeverio, S.; Gesztesy, F.; Hoegh-Krohn, R.; Holden, H. \emph{Solvable Models in Quantum Mechanics}; Springer Science \& Business Media: Berlin/Heidelberg, Germany, 2012.

\bibitem{Measure1}
Halmos, P.R.  \emph{Measure Theory};  Springer: Berlin/Heidelberg, Germany, 2013; Volume 18.
\bibitem{Measure2}
Tao, T. (Ed.)  \emph{An Introduction to Measure Theory}; American Mathematical Association: Washington, DC, USA, 2011; Volume 126.
\bibitem{Measure3}
Swartz, C.W.  \emph{Measure, Integration and Function Spaces}; World Scientific: Singapore, 1994.
\bibitem{Measure4}
Bartle, R.G. \emph{The Elements of Integration and Lebesgue Measure}; John Wiley Sons: Hoboken, NJ, USA, 2014.

\bibitem{Gomis1}
Jaume, G.;  Mehen, T. Space–time noncommutative field theories and unitarity. \emph{Nucl. Phys. } \textbf{2000}, \emph{591}, 265--276.
\bibitem{Gomis2}
\textls[-15]{Bahns, D.  Unitary quantum field theory on the noncommutative Minkowski space. \emph{Fortschritte Der Phys. Prog. Phys.} \textbf{2003}, \emph{51}, 658--663.}
\bibitem{Gomis3}
Morita, K.; Okumura, Y.; Umezawa, E. Lorentz invariance and the unitarity problem in non-commutative field theory. \emph{Prog. Theor. Phys.} \textbf{2003}, \emph{110}, 989--1001. [\href{http://dx.doi.org/10.1002/prop.200310079}{CrossRef}]

\bibitem{Rychkov1}
Hogervorst, M.; Rychkov, S.; van Rees, B.C.  Unitarity violation at the Wilson-Fisher fixed point in 4-$\epsilon$ dimensions. \emph{Phys. Rev. D} \textbf{2016}, \emph{93}, 125025. [\href{http://dx.doi.org/10.1143/PTP.110.989}{CrossRef}]
\bibitem{Rychkov2}
\textls[-35]{Jin, Q.; Ren, K.; Yang, G.; Yu, R.  Is Yang-Mills Theory Unitary in Fractional Spacetime Dimensions? \emph{arXiv} \textbf{2023}, arXiv:2301.01786. [\href{http://dx.doi.org/10.1103/PhysRevD.93.125025}{CrossRef}]}

\bibitem{open1}
Rivas, A.; Huelga, S.F.  \emph{Open Quantum Systems}; Springer: Berlin/Heidelberg, Germany, 2012; Volume 10, pp. 978--983 [\href{http://dx.doi.org/10.1007/s11433-024-2370-6}{CrossRef}]
\bibitem{open2}
Breuer, H.P.; Petruccione, F.  \emph{The Theory of Open Quantum Systems}; Oxford University Press: Oxford, MS, USA,  2002.
\bibitem{open3}
Rotter, I.; Bird, J.P.  A review of progress in the physics of open quantum systems: Theory and experiment. \emph{Rep. Prog. Phys.} \textbf{2015}, \emph{78}, 114001.
\bibitem{open4}
Isar, A.; Sandulescu, A.; Scutaru, H.; Stefanescu, E.; Scheid, W. Open quantum systems. \emph{Int. J. Mod. Phys. E} \textbf{1994}, \emph{3},  635--714. [\href{http://dx.doi.org/10.1088/0034-4885/78/11/114001}{CrossRef}] [\href{http://www.ncbi.nlm.nih.gov/pubmed/26510115}{PubMed}]

\bibitem{jimwlk}
Lublinsky, M.; Mulian, Y. High Energy QCD at NLO: From light-cone wave function to JIMWLK evolution. \emph{J. High Energy Phys.} \textbf{2017}, \emph{2017}, 97. [\href{http://dx.doi.org/10.1142/S0218301394000164}{CrossRef}]

\bibitem{iancu1}
Iancu, E.; Mulian, Y. Forward trijet production in proton–nucleus collisions. \emph{Nucl. Phys. A} \textbf{2019}, \emph{985}, 66--127. [\href{http://dx.doi.org/10.1007/JHEP05(2017)097}{CrossRef}]
\bibitem{iancu2}
Iancu, E.; Mulian, Y. Forward dijets in proton-nucleus collisions at next-to-leading order: The real corrections. \emph{J. High Energy Phys.} \textbf{2021}, \emph{2021}, 5. [\href{http://dx.doi.org/10.1016/j.nuclphysa.2019.02.003}{CrossRef}]

\bibitem{adhoc1}
Chen, Z.; Mueller, A.H. The dipole picture of high energy scattering, the BFKL equation and many gluon compound states. \emph{Nucl. Phys. B} \textbf{1995}, \emph{451}, 579--604. [\href{http://dx.doi.org/10.1007/JHEP03(2021)005}{CrossRef}]
\bibitem{adhoc2}
\textls[-15]{Marquet, C. Forward inclusive dijet production and azimuthal correlations in pA collisions. \emph{Nucl. Phys. A} \textbf{2007}, \emph{796}, 41--60. [\href{http://dx.doi.org/10.1016/0550-3213(95)00350-2}{CrossRef}]}

\bibitem{Dumitru}
Adrian, D.; Paatelainen, R. Sub-femtometer scale color charge fluctuations in a proton made of three quarks and a gluon. \emph{Phys. Rev. D} \textbf{2021}, \emph{103},  034026. [\href{http://dx.doi.org/10.1016/j.nuclphysa.2007.09.001}{CrossRef}]

\bibitem{Lindelof}
Lindelöf, E.  Sur l'application de la méthode des approximations successives aux équations différentielles ordinaires du premier ordre. \emph{Comptes Rendus Hebd. Des SéAnces L'AcadéMie Des Sci.} \textbf{1894}, \emph{118}, 454--457.

\bibitem{Evolutiona}
Seifert, C.; Trostorff, S.; Waurick, M.  \emph{Evolutionary Equations: Picard's Theorem for Partial Differential Equations, and Applications}; Springer Nature: Berlin/Heidelberg, Germany, 2022; p. 317.

\bibitem{lax}
Lax, P.D.;  Richtmyer, R.D.  Survey of the stability of linear finite difference equations. \emph{Commun. Pure Appl. Math.} \textbf{1956}, \emph{9}, 267--293.

\bibitem{byron}
\textls[-15]{Byron, F.W.; Fuller, R.W. \emph{Mathematics of Classical and Quantum Physics}; Courier Corporation: Chelmsford, MA, USA, 2012. [\href{http://dx.doi.org/10.1002/cpa.3160090206}{CrossRef}]}

\bibitem{distrib1}
Lighthill, M.J. \emph{An Introduction to Fourier Analysis and Generalised Functions}; Cambridge University Press: Cambridge, UK, 1958.
\bibitem{distrib2}
Ferreira, J.C.; Hoskins, R.F.; Sousa-Pinto, J. \emph{Introduction to the Theory of Distributions}; Cambridge University Press: Cambridge, UK,~1998.
\bibitem{distrib3}
Hoskins, R.F. \emph{Delta Functions, Introduction to Generalised Functions}; Woodhead Publishing: Chichester, UK, 2009.

\bibitem{Schwarz}
Schwartz, L.  Sur l’impossibilité de la multiplication des distributions. \emph{CR Acad. Sci. Paris} \textbf{1954}, \emph{239}, 6.

\bibitem{Kan1}
Kanwal, R.P.  \emph{Generalized Functions: Theory and Applications}; Springer Science \& Business Media: Berlin/Heidelberg, Germany,~2004.
\bibitem{Kan2}
Estrada, R.; Kanwal, R.P. \emph{Singular Integral Equations}; Springer Science \& Business Media: Berlin/Heidelberg, Germany, 2000.

\bibitem{Sylvester1}
Sylvester, J.  Sur l'equations en matrices $px=xq$. \emph{C. R. Acad. Sci. Paris} \textbf{1884}, \emph{99}, 67--71. 115--116.

\bibitem{Sylvester2}
Bartels, R.H.; Stewart, G.W.  Solution of the matrix equation $AX+XB=C$. \emph{Comm. ACM} \textbf{1972}, \emph{15}, 820--826. [\href{https://doi.org/10.1145/361573.361582}{CrossRef}]

\bibitem{Lyapunov}
Parks, P.C.  AM Lyapunov's stability theory---100 years on. \emph{IMA J. Math. Control. Inf.} \textbf{1992}, \emph{9}, 275--303. [\href{http://dx.doi.org/10.1145/361573.361582}{CrossRef}]
\end{thebibliography}
\end{document}